\documentclass[intlimits,twoside,a4paper]{article}
\usepackage{amsmath,amssymb,gensymb}
\usepackage{graphicx}

\usepackage[cp1251]{inputenc}

\usepackage{cmpj3}

\usepackage{bm}

\newcommand{\ps}{\,\textrm{ps}}
\newcommand{\fs}{\,\textrm{fs}}

\issue{2017}{20}{2}{23003}
\doinumber{10.5488/CMP.20.23003}

\title[Curcumin in solvents]
{On the properties of a single OPLS-UA model curcumin molecule
in water, methanol and dimethyl sulfoxide.
Molecular dynamics computer simulation results}

\author[T. Patsahan, J.M. Ilnytskyi, O. Pizio]{T. Patsahan\refaddr{label1}, J.M. Ilnytskyi\refaddr{label1},
        O. Pizio\refaddr{label2}\thanks{Corresponding author, E-mail:~oapizio@gmail.com. On sabbatical leave from Instituto de Qu\'{i}mica de la UNAM. }}
\addresses{
\addr{label1} Institute for Condensed Matter Physics of the National Academy of Sciences of Ukraine, \\ 1 Svientsitskii St., 79011 Lviv, Ukraine
\addr{label2} Instituto de Investigaciones en Materiales, Universidad Nacional Aut\'{o}noma de M\'{e}xico,
Circuito Exterior, 04510, Cd. Mx., M\'{e}xico}

\date{Received March 17, 2017, in final form May 4, 2017}

\begin{document}

\maketitle

\begin{abstract}
The properties of model solutions consisting of a solute --- single curcumin molecule
in water, methanol and dimethyl sulfoxide solvents
have been studied using molecular dynamics (MD) computer simulations
in the isobaric-isothermal ensemble.
The united atom OPLS force field (OPLS-UA) model for curcumin molecule
proposed by us recently [\,J. Mol. Liq., 2016, \textbf{223}, 707]
in combination with the SPC/E water, and the OPLS-UA type models for methanol and dimethyl sulfoxide
have been applied.
We have described changes of the internal structure of the solute molecule
induced by different solvent media in very detail.
The pair distribution functions between particular fragments of a solute molecule with solvent particles
have been analyzed.
Statistical features  of the hydrogen bonding between different species were explored.
Finally, we have obtained a self-diffusion coefficient of curcumin molecules in three model solvents.

\pacs 02.70.Ns, 61.20.Ja, 82.30.Rs, 87.15.hp
\end{abstract}

\section{Introduction}

Curcumin, derived from the root of turmeric {\it Curcuma longa}, is well known not only as
a spice and natural colorant but also as an important substance in biomedicine due to
possible ample  applications.
Namely, the issues concerning the
antioxidant, anti-inflammatory, antiviral and anticancer activity of curcumin have become the subject of numerous
experimental studies for the recent few decades, see e.g.,
recent reviews and references therein~\cite{ghosh1,kumar1,luthra1,mehanny1,ngo1}.
We are unable to comment on the problems related to the applications of curcumin-based
agents in medicinal chemistry~\cite{nelson} and just wish to mention that
experimental research has urged the use of methods of theoretical chemistry in this area,
more specifically, methods of quantum chemistry. However, as noted by Wright~\cite{wright1}, from a theoretical modelling
perspective, there is a bewildering number of variables involved in the experiments.
Consequently, there is much room for the development of adequate modelling and
for confronting predictions from computer simulations against experimental data.

The studies of liquid systems containing curcumin solutes by using computer simulation techniques
have been initiated quite recently
\cite{ngo2,Suhartanto-2012,Varghese-2009,Wallace-2013,Samanta-2013,Hazra-2014,Sreenivasan-2014,Yadav-2014,Parameswari-2015,Priyadarsini-2009}.
The works~\cite{Samanta-2013,Hazra-2014} are of particular interest for the present study. In each of them,
the modelling of a curcumin molecule force field has been undertaken starting from quantum chemical
calculations within the B3LYP DFT method by using different versions of Gaussian software. Unfortunately,
not all the details of modelling were presented in the original papers and in supplementary
materials to them. For, example, in \cite{Hazra-2014} only charges for the interaction sites are given
whereas all parameters concerning bonds, angles and dihedral angles were omitted.
This was one of the reasons that in the previous work from this laboratory we have developed
the OPLS-united atom model for enol-tautomer of curcumin molecule and tested
it in vacuum and water using classical MD  simulations~\cite{Ilny-2016}.
It has been already mentioned \cite{wright1} that
experiments were performed in a variety of solvents (e.g., chlorobenzene,
isopropanol-water mixture, dimethylsulfoxide, acetonitrile) due to marginal  solubility of curcumin
in pure water.
Therefore, in the current study we extend our recent work by considering a single
curcumin molecule in pure methanol (MeOH) and pure dimethylsulfoxide (DMSO).
It is worth mentioning that specific features of curcumin molecule conformations
in liquid DMSO were studied experimentally quite recently~\cite{Slabber-2016}.
Our principal objective here is to elucidate the similarities and differences in the behavior of this
molecule in water, MeOH and DMSO. With this aim we also needed to recalculate some of the results
for curcumin in water from the previous study~\cite{Ilny-2016} in order to make the time extent of
computer simulations for all three systems  similar.

\section{Model and technical details of calculations}

Simulations of a single curcumin molecule in three solvents, H$_2$O, MeOH and DMSO, were performed
in the isothermal-isobaric (NPT) ensemble at a temperature of $298.15$~K and at $1$~bar.
In all cases we used GROMACS simulation engine~\cite{GROMACS} version 4.6.7.

\begin{figure}[!b]
\begin{center}
\includegraphics[width=90mm,angle=0,clip=true]{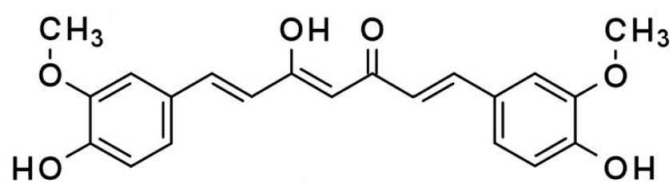}
\caption{\label{fig_enol} Chemical structure of the enol form of curcumin.}
\end{center}
\end{figure}

\begin{figure}[!b]
\begin{center}
\includegraphics[width=90mm,angle=0,clip=true]{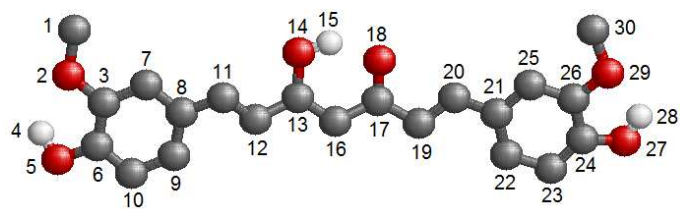}
\caption{\label{fig_curc} (Color online) Schematic representation for the united-atom
curcumin model with sites numbering. Carbon groups are shown as dark gray spheres,
oxygens --- as red spheres, hydrogens --- as small light-gray spheres.}
\end{center}
\end{figure}

In close similarity to the previous work~\cite{Ilny-2016}, our attention is restricted to the
 enol-tautomer of curcumin solely. Its chemical formula is given in figure~\ref{fig_enol}.
This tautomer was investigated  in the recent computer simulation
studies~\cite{Samanta-2013,Hazra-2014}. It is a dominant tautomer both in
solids and in various solutions
as confirmed  in various works,
see e.g., \cite{Slabber-2016,Kawano-2013,Kolev-2005,Cornago-2008}.
Besides, it is the major phytoconstituent of extracts
of {\it Curcuma longa}  according to \cite{nelson}.
Thus, we are interested in the effects of solvents
on a particular tautomer, while diketo-tautomer will be explored elsewhere.
A ``ball-and-stick'' schematic representation of the curcumin molecule is shown
in figure~\ref{fig_curc}, with labels denoting the sites (carbon groups, oxygen and hydrogen) of the
united atom force field model.
All the parameters of the force field are taken from~\cite{Ilny-2016}, see supporting information file
to that article.

For water we used the SPC/E model~\cite{SPCE_model}.
For MeOH, the OPLS-UA model was taken according to~\cite{Leeuwen}.
However, in contrast to \cite{Leeuwen}, we used the harmonic bond-angle potential for CH$_3$-O-H with
the force constant from the OPLS-UA database~\cite{OPLS-1996}.
In the case of DMSO, we used the four-site model with all parameters taken from the OPLS-UA database,
except that the improper dihedral angle S-CH$_3$-O-CH$_3$ was added, it was described by the
GROMOS set of parameters~\cite{Oostenbrink}.

The geometric combination rules were used to determine parameters for cross interactions (rule 3
of GROMACS package). To evaluate the contributions of Coulomb interactions the particle mesh Ewald method was used
(fourth-order spline interpolation and grid spacing for fast Fourier transform equal to $0.12$).
The electrostatic and Lennard-Jones cut-off distances were chosen equal to $1.1$~nm.
The van der Waals tail correction terms to the energy and pressure were taken into account.
The lengths of the bonds were constrained by LINCS.
For each system, a periodic cubic simulation box was set up with $2000$ solvent particles and
a single curcumin molecule.

The initial configuration of particles was prepared by placing them randomly in the simulation box
using the GROMACS genbox tools. Next, the systems underwent energy minimization to remove the
possible overlap of atoms in the starting configuration. This was done by applying the steepest descent algorithm.
After minimization we performed equilibration of each system for $50$~ps at
a temperature of $298.15$~K and at $1$~bar
using a timestep $0.1$~fs.

The Berendsen thermostat with the time coupling constant of $0.1\ps$ and isotropic Berendsen barostat
with the time constant of $2$~ps were used. We are aware that this thermostat
may yield certain inaccuracies in common MD, see, e.g., comments in \cite{Rosta-2009}
concerning the application of replica exchange MD to conformational equilibria, and becomes completely unsatisfactory
in calculations of fluctuation-based  properties. Still, for purposes of the present work, and
to keep a consistency with the previous development~\cite{Ilny-2016}, we have decided to use
this thermostat for the moment.

The compressibility parameter was taken equal to
$4.5\cdot 10^{-5}$~bar$^{-1}$ for water, $1.2\cdot 10^{-4}$~bar$^{-1}$ for
methanol, similar to \cite{roccatano2}, and $5.25\cdot 10^{-5}$~bar$^{-1}$ for DMSO.
All the properties were collected by performing the averages
over $10$ consecutive simulation runs of the length of $50$~ns.
Each run started from the last configuration of the previous one.
The timestep in the production runs was chosen to be $1\fs$.

\section{Results and discussion}

First, we would like to discuss the changes of the internal structure of a single curcumin molecule
induced by each solvent under study. Various properties  are described in terms of histograms of the
probability distributions termed as the distribution functions.
For each production run, we select a set of frames with the distance
in time equal to 2~ps.
The outputs for any property of interest are collected on a set of frames and the resulting
histogram is then normalized. Next, the average over a set of runs is performed.

\subsection{Distributions of dihedral angles }
\begin{figure}[!t]
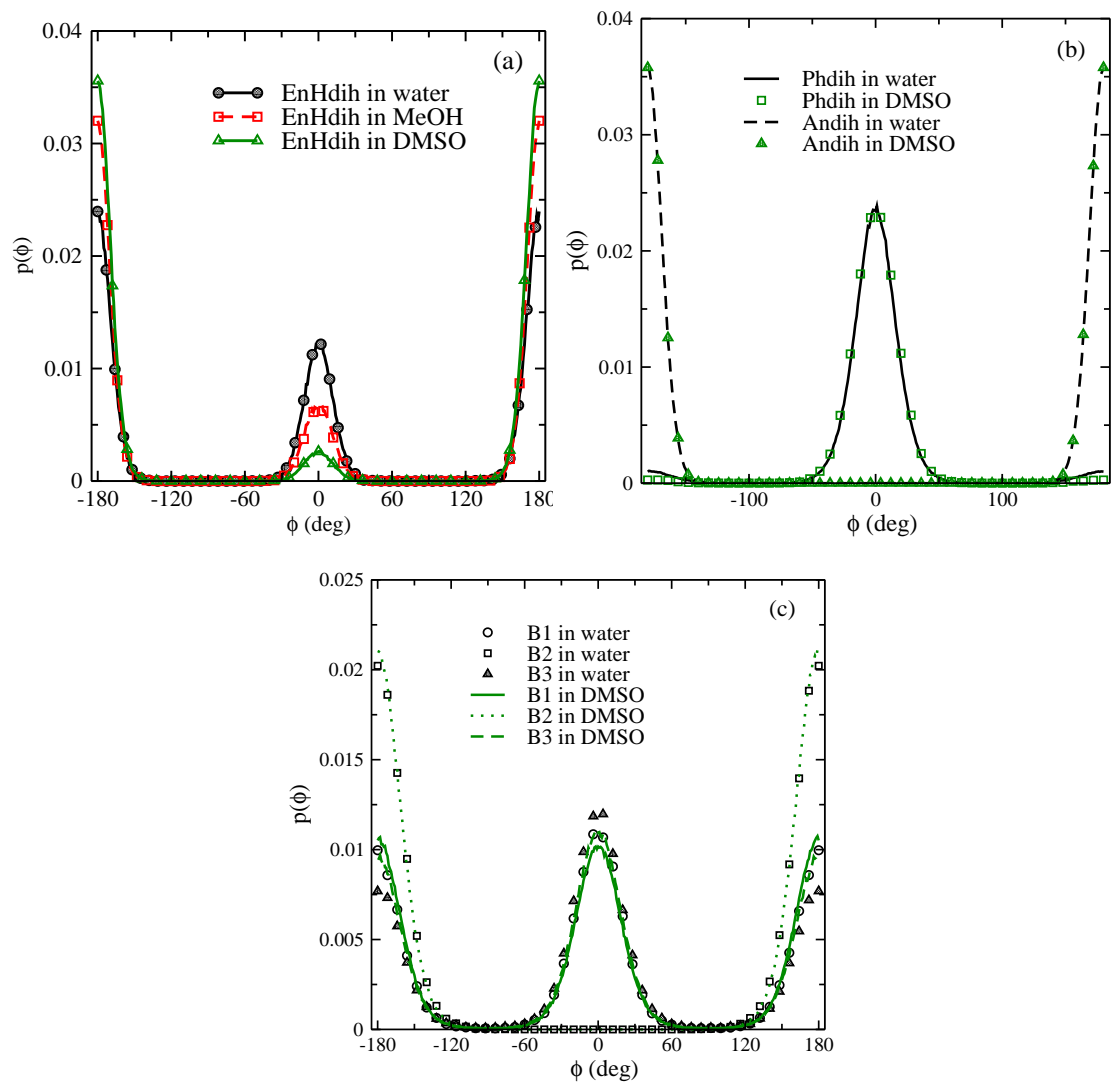

\begin{center}
\includegraphics[width=0.48\textwidth]{distr_dih1}
\includegraphics[width=0.48\textwidth]{distr_dih2}\\
\vspace{4mm}
\includegraphics[width=0.48\textwidth]{distr_dih3}
\caption{\label{fig_dihedral} (Color online) Histograms of the probability distributions for the dihedral angles of curcumin
molecule in different solvents.}
\end{center}
\end{figure}

The histograms of  probability distributions of the dihedral angles of curcumin in water, MeOH and DMSO
are presented in three panels of figure~\ref{fig_dihedral}.
The abbreviated notations for dihedral angles are given in table~\ref{tab_set_dih}.

\begin{table}[!h]
  \centering
   \caption{\label{tab_set_dih} Nomenclature for the groups
 of dihedral angles of a curcumin molecule. Site numbers are from figure~\ref{fig_curc}.}
 \vspace{2ex}
  \begin{tabular}{|c|ccc|}
  \hline\hline
  dih. group abbr.  & \multicolumn{3}{c|}{dihedrals}\\
  \hline\hline
  $\textrm{EnHdih}$ & \multicolumn{3}{c|}{12-13-14-15}\\
  $\textrm{Phdih}$  & 3-6-5-4     & and & 26-24-27-28\\
  $\textrm{Andih}$  & 6-3-2-1     & and & 24-26-29-30\\
  $\textrm{B}_1$  & 12-11-8-7   & and & 25-21-20-19\\
  $\textrm{B}_2$  & 13-12-11-8  & and & 21-20-19-17\\
  $\textrm{B}_3$  & 16-13-12-11 & and & 20-19-17-16\\
  \hline\hline
 \end{tabular}
\end{table}

To begin with, we consider the EnHdih dihedral angle distribution [figure~\ref{fig_dihedral}~(a)]
that describes the orientation of hydrogen atom H15 within the enol group.
For all three solvents in question, $p(\phi)$ is characterized by the maxima
at $\phi=0\degree$ and $\phi=\pm 180\degree$ of different height.
The maxima located at $\phi=\pm 180\degree$ indicate that H15 is preferentially directed
toward the ketone oxygen (O18), rather than outward the enol group.
Thus, the trans-conformation of EnHdih angle is dominant compared to the cis-conformation in all cases.
It seems that intramolecular electrostatic interaction between  H15 and O18 leads to these trends.
However, solvent effect is manifested in the difference of heights of the maxima at characteristic angles.
Apparently, the cis-conformation is suppressed, comparing to its trans-counterpart in two solvents capable of solubilizing
curcumin. By contrast, in water, H15 has enough freedom to accommodate in both conformations. An abundant
population of cis-conformation witnesses a possibility of the formation of H-bonds between water oxygens and
H15 belonging to the enol group.
It is worth noting that for curcumin molecule in vacuum, the only state of EnHdih is trans-conformation~\cite{Ilny-2016}.

Histograms of probability distribution of dihedral angles of phenol and anisole groups
[figure~\ref{fig_dihedral}~(b)] practically coincide for three systems studied [the curve describing the influence of MeOH on $p(\phi)$ is omitted to make the figure less loaded].
Thus,  interpretation of the results is similar for all cases.
The Phdih dihedrals have a principal maximum at $\phi=0\degree$ (cis-conformation)
due to the hydrogens H4 and H28 directed toward the oxygens O2 and O29, respectively.
Trans-conformation for Phdih can happen, but with negligible probability as it follows from the figure.
The Andih angle distributions have solely two maxima at $\phi=\pm 180\degree$.
In the trans-conformation,  the methyl groups, C1 and C30, point away from the oxygens O5 and O27, respectively.

Dihedral distributions, $\mathrm{B_1}$, $\mathrm{B_2}$ and $\mathrm{B_3}$ [figure~\ref{fig_dihedral}~(c)] characterize the flexibility of the spacer connecting aromatic rings to the central enol group.
The changes in the $\mathrm{B_1}$ dihedrals describe the rotation of aromatic rings apart from the spacer,
whereas the changes of the $\mathrm{B_2}$ and $\mathrm{B_3}$ are related to the rotation of all the side branches
including aromatic rings and the spacer fragments connected to them.
The $\mathrm{B_2}$ distribution qualitatively differs
from the $\mathrm{B_1}$ and $\mathrm{B_3}$ curves.
While the $\mathrm{B_1}$ and $\mathrm{B_3}$ histograms have maxima at $\phi=0\degree$ and $\phi=\pm 180\degree$,
the $\mathrm{B_2}$ has a maximum around $\phi=\pm 180\degree$ (trans-conformation).
Consequently, there is no rotation around the $11{-}12$ and $19{-}20$ bonds.
The same behaviour for the $\mathrm{B_2}$ is observed for a curcumin molecule in water and in DMSO,
two curves practically coincide (the MeOH curve is omitted).
The difference between the $\mathrm{B_1}$ and $\mathrm{B_3}$ curves
in water and in DMSO is better pronounced but small.
We are unaware of any direct experimental spectroscopic evidence or computer simulation studies of the
relevant properties to estimate the quality of our predictions. Nevertheless, some indirect supporting arguments
are given in the following subsection.

\subsection{Distance and angles distributions}

\begin{figure}[!b]
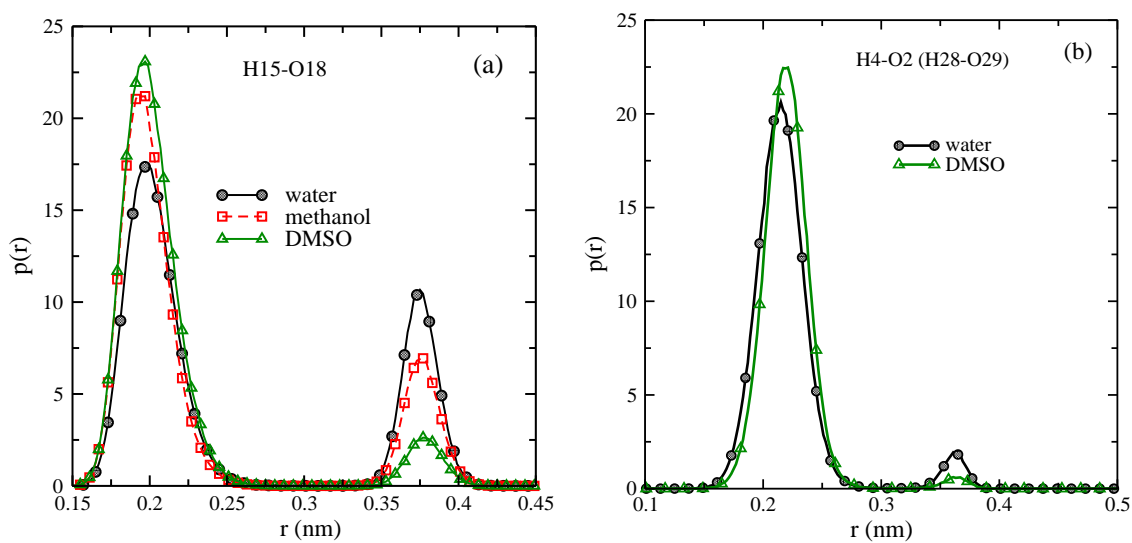

\begin{center}
\includegraphics[width=0.48\textwidth]{distr_dist3} \hspace{2mm}
\includegraphics[width=0.48\textwidth]{dist_OHphen}
\caption{\label{fig_distance} (Color online)
Histograms of the probability distribution of the H15-O18~(a) and H4-O2 (H28-O29)~(b) distances for
curcumin in different solvents.}
\end{center}
\end{figure}

The distribution function of the EnHdih angle discussed in the
previous subsection can be interpreted in terms of the changes
of the distance between hydrogen, H15, and ketone oxygen, O18,  figure~\ref{fig_distance}~(a).
Well pronounced maxima at  $0.196$~nm and $0.375$~nm
correspond to  H15 pointing toward and outward the enol group, respectively.
As expected from the behavior of EnHdih [figure~\ref{fig_dihedral}~(a)],
the site H15 prefers to be directed toward the enol group in all solvents considered.
The maximum of the O18-H15 distance distribution at $0.196$~nm is higher for the system with DMSO solvent,
in comparison with MeOH and water. On the other hand, the maximum at $0.375$~nm is much lower in magnitude
for curcumin-DMSO system compared with two other solvents.
Consequently, the trans-conformation of the enol group is most frequent and
possibly most stable in thermodynamic sense for curcumin-DMSO solution.

This is a valuable result. There is no experimental evidence supporting our observation. However, in
computer simulations of Samanta and Roccatano~\cite{Samanta-2013}, the H15-O18 distance distribution
was discussed. According to their  model~\cite{Samanta-2013},
the H15-O18 distance distribution in water solvent is bimodal with two maxima
of almost equal height located at $\approx 0.16$ and $0.35$~nm.
These values do not differ much from what we observe.
It seems, however, that the discrepancy of the absolute values of maxima
reported in figure~4 of~\cite{Samanta-2013} and our data in figure~\ref{fig_distance}~(a)
is due to the differences of parameters used in modelling the dihedral angles.
Moreover, in the case of MeOH solvent,
the second maximum is totally absent in the model of \cite{Samanta-2013}
whereas in our distribution it is seen.

The trends of the behavior of the  distances H4-O2 and H28-O29 have been analyzed by us as well.
Histograms of the  probability distribution for each of these distances
in figure~\ref{fig_distance}~(b) confirm our conclusions concerning the
dihedral angles Phdih [cf. figure~\ref{fig_dihedral}~(b)].
In particular, one can observe a bimodal distribution with two maxima
at $r=0.22$~nm and $0.36$~nm. They correspond to the cis- and
trans-conformations. It is clearly seen that the larger distance (trans-conformation) is realized
with very low probability in water and DMSO solvents, though in DMSO it is noticeably lower.
Therefore, we can conclude that such a rare change of Phdih conformation can hardly have an effect
on the overall behavior of the curcumin molecule.

\begin{figure}[!b]
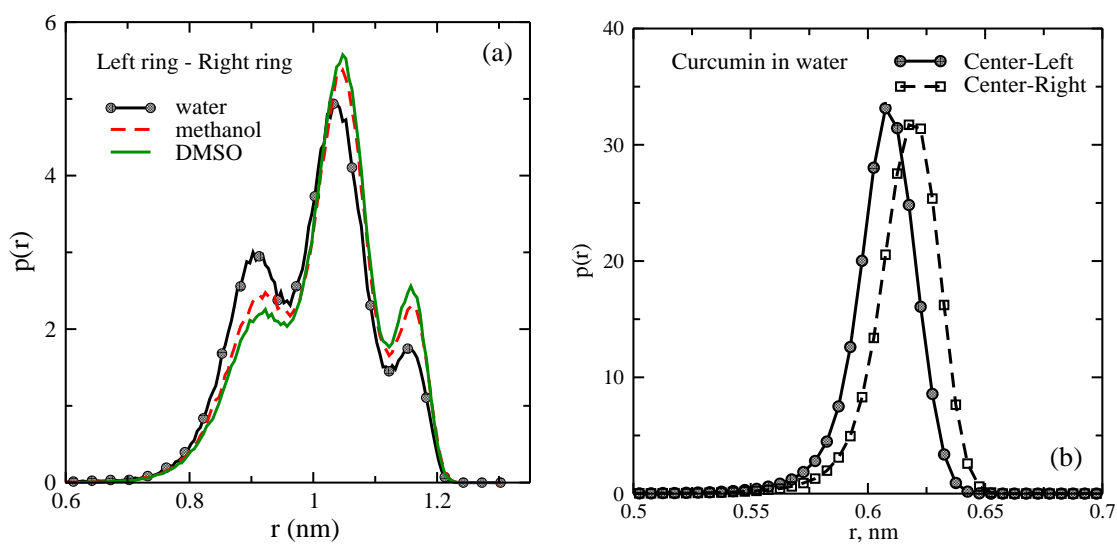

\vspace{2ex}
\begin{center}
\includegraphics[width=0.46\textwidth]{distr_dist1} \hspace{3mm}
\includegraphics[width=0.48\textwidth]{distr_dist2}
\caption{\label{fig_ring_distance} (Color online)
Histograms of the probability distribution of the ring-ring~(a) and center-ring~(b) distances for curcumin in different solvents.}
\end{center}
\end{figure}

Probability distributions describing the flexibility of a curcumin molecule
in three solvents are shown in figure~\ref{fig_ring_distance}~(a) and figure~\ref{fig_ring_distance}~(b).
The Left ring-Right ring distance is defined as a distance between centers of aromatic rings,
whereas the Center-Left and Center-Right are the distances from the center of the enol ring to the center
of the left and  right ring, respectively.
The center of a  ring is chosen as a center of mass of carbons and hydrogens composing it.
For the Left-Right distance of curcumin in water presented in figure~\ref{fig_ring_distance}~(a),
three well pronounced maxima at  $0.90$~nm, $1.03$~nm and $1.15$~nm are observed.

As it was discussed in our previous study~\cite{Ilny-2016}, this shape is explained by the conformational states
available for the $\mathrm{B_1}$, $\mathrm{B_2}$ and $\mathrm{B_3}$ dihedrals.
It was also concluded that only the $\mathrm{B_3}$ dihedrals involving
 16-13-12-11 and 20-19-17-16 groups are responsible for the Left-Right
distance of a curcumin molecule. Since only three possible combinations of cis- and trans-
conformation states of these two groups may occur,
the number of the most probable Left-Right distances is three.

The maximum which corresponds to the middle distance $1.03$~nm is the
most pronounced, and consequently this structure is the most probable, figure~\ref{fig_ring_distance}~(a).
The left-hand maximum in the figure corresponding to the distance $0.90$~nm is much lower.
Finally, the lowest maximum can be found at the distance $1.15$~nm.
The obtained distances agree well with the results from our previous investigation of curcumin in
SPC-E water~\cite{Ilny-2016}.

In the case of the MeOH solvent [figure~\ref{fig_ring_distance}~(a)], we again observe three maxima,
but of different heights, in comparison with water.
Moreover, all of them are slightly shifted towards higher distances.
The middle peak becomes higher in a ``better''  solvent that solubilizes curcumin in contrast to water.
Specifically, in the case of MeOH, the middle maximum of the Left-Right distance distribution is higher
than in  water. It is even higher for curcumin in DMSO compared to the curcumin
molecule in MeOH.  Similar behavior  is observed for the right-hand maximum.
By contrast,  opposite trends are seen for the left-hand maximum.
Finally, for curcumin molecules in  DMSO, the height of the right-hand maximum becomes higher than the left-hand one.

The Center-Left and Center-Right distance distributions are shown in figure~\ref{fig_ring_distance}~(b)
only for the curcumin in water, since for MeOH and DMSO,  the results are very close
to the system with water. As can be seen, both  distributions have a single  maximum,
approximately of equal heights, just slightly shifted.
The Center-Left distance has the most probable value at $0.61$~nm whereas the Center-Right most
probable distance is slightly higher. The shift can be attributed to the asymmetry of the spacer
of the curcumin molecule.
We are not aware of the data coming from other models of a curcumin molecule in various solvents. However, a detailed
discussion of fine features of the molecular internal structure  can be helpful in designing more
sophisticated  models in future research. Namely, the modification and
reparameterization of the dihedral-angle potentials can be attempted along the lines proposed by
Kurokawa et al.~\cite{Kurokawa-2012}.

Changes of the internal structure of a molecule  can be interpreted in terms of the angles
between certain fragments of the curcumin molecule immersed in a solvent.
To begin with,  we explored the trends of the behavior of the angle between planes of aromatic rings.
A plane is determined by picking up three sites belonging to a ring,
namely C6, C7, C9 for the left-hand ring and C22, C24, C25 for the right-hand ring.
The distribution of the angles between the rings, $\alpha$,
has been analysed, but the figure is not shown for the sake of brevity.
The probability distributions of $\alpha$ are very similar
in all the solvents considered and they
almost coincide with the histogram for $p(\alpha)$ presented in figure~10 of \cite{Ilny-2016}.
The $p(\alpha)$ distribution has a bimodal shape principally determined by the intrinsic structure
of a molecule within the framework of the modelling applied.

\begin{figure}[!b]
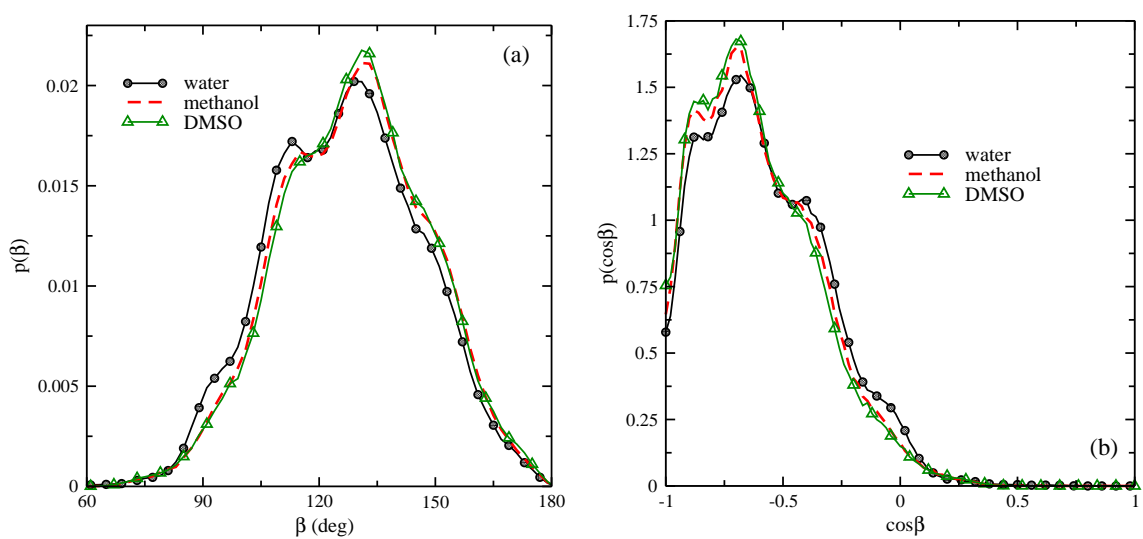

\begin{center}
\includegraphics[width=0.49\textwidth]{distr_ang2} \hspace{2mm}
\includegraphics[width=0.48\textwidth]{distr_ang2cos}
\caption{\label{fig_angle} (Color online) Histograms of the probability distributions of angles
$\beta$ (defined in the text) and angle cosine $\cos(\beta)$, in panels (a) and (b), respectively,
for the curcumin molecule in different solvents.}
\end{center}
\end{figure}

Another characteristic angle is $\beta$, which is the side-side angle defined via the triplet
of sites, C3-C16-C26. Its behavior is given in terms of
angular distribution $p(\beta)$, figure~\ref{fig_angle}~(a).
Actually, the  $p(\beta)$ distribution is unimodal with the maximum at
$\approx 130\degree$. At a larger value of the angle, there is a shoulder for all three solvents.
However, the shape of $p(\beta)$
in DMSO is very close to that in MeOH. Only in water, a weakly pronounced secondary maximum
can be observed at a smaller angle, $\approx 112\degree$.
These distributions
reflect the bending of curcumin molecules and in some sense should yield information
similar to figure~\ref{fig_ring_distance}~(a). However, splitting into three maxima
using the distance variable [figure~\ref{fig_ring_distance}~(a)] in terms of the angles is manifested
solely in shoulders of the $p(\beta)$ distribution.
Nevertheless, the angular distribution curves
indicate that the curcumin molecule is slightly more bent in water compared to DMSO or MeOH.
In figure~\ref{fig_angle}~(b),  we also show the corresponding distributions for
the angle cosine [$\cos(\beta)$].
It provides conclusions similar to $p(\beta)$,
though in such representation the shoulders become more prominent.

\subsection{Dipole moment}

\begin{figure}[!b]
\begin{center}
\includegraphics[width=75mm,angle=0,clip=true]{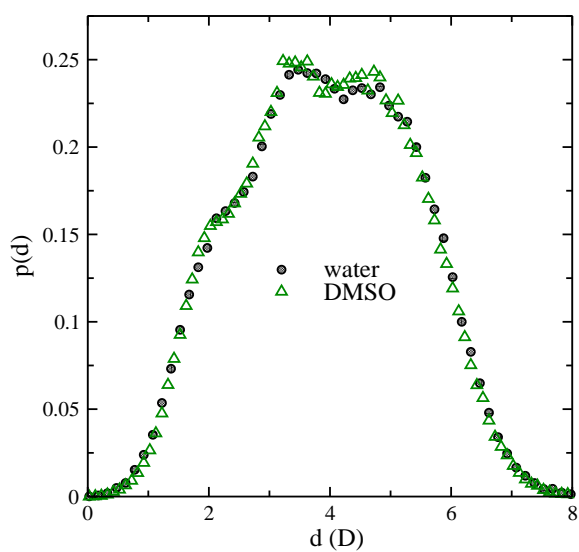}
\caption{\label{fig_dipole} (Color online) A comparison of the probability distribution for the dipole moment $d$ of
the curcumin molecule in water and DMSO.}
\end{center}
\end{figure}

The  dipole moment distribution only marginally changes upon the solvent change
within the framework of the model.
Therefore, only the histograms of the probability distribution of the  magnitude of the
dipole moment of a curcumin molecule in water and DMSO are presented in figure~\ref{fig_dipole},
the curve for MeOH is omitted.

\begin{figure}[!t]
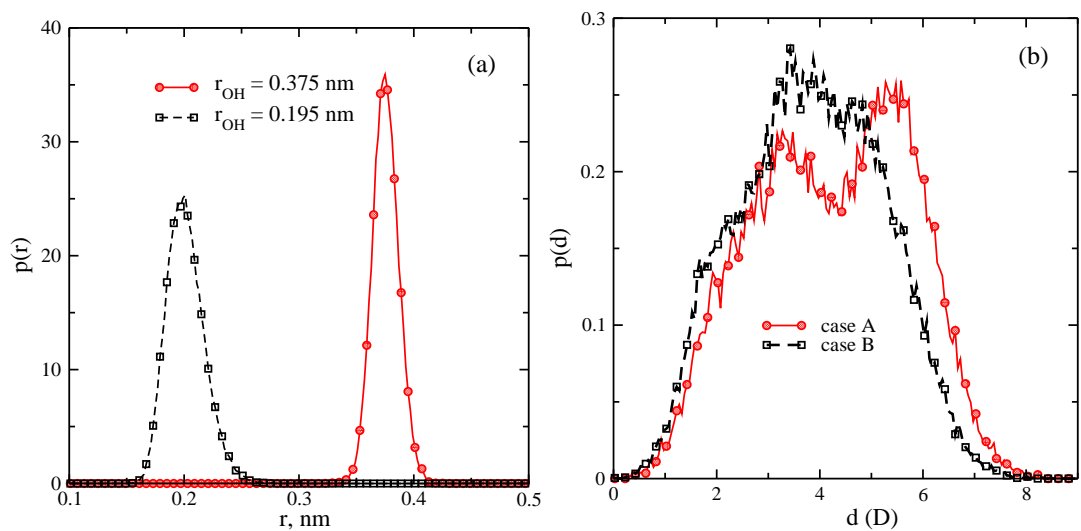

\begin{center}
\includegraphics[width=70mm,angle=0,clip=true]{dip_1_2a}
\includegraphics[width=70mm,angle=0,clip=true]{dip_1_2b}
\caption{\label{fig_dipole2} (Color online)
Histograms of the probability distribution of O14-H15 distance $p(r)$
obtained from different fragments of curcumin trajectory, cases A and B, left-hand panel.
The corresponding probability distribution for the dipole moment $p(d)$ of
curcumin in water  obtained by using these trajectories (right-hand panel).}
\end{center}
\end{figure}

The dipole moment distribution reflects
superposition of the conformational states of a curcumin molecule discussed in terms
of dihedral angles and the related properties in the previous subsections.
It seems that an essential conformation change is the position of hydrogen site
in the enol group. The splitted maximum in the dipole moment distribution $p(d)$
is due to the enol hydrogen, H15, pointing toward ketone-oxygen, O18, or
outward the enol group.
To get a profounder insight,  we have analyzed two distributions obtained
from the curcumin trajectories. Namely, we picked up a part of the trajectory  when H15
is pointing solely outwards (case A: $r_{\text{OH}}=0.375$~nm) and the other part
when H15 is directed  toward (case B: $r_{\text{OH}}=0.195$~nm)
the enol group. Histograms describing $p(r)$
for these trajectories are given in figure~\ref{fig_dipole2}~(a).
The corresponding $p(d)$ functions are presented in figure~\ref{fig_dipole2}~(b).
The $p(r)$ is clearly a unimodal function in each case. However,
in the case~A, we observe two maxima $d=3.5$~D and $d=5.3$~D on $p(d)$
coming from distinct conformation states. Additional exploration leads to
the conclusion that this bimodality in the present case comes from the
behavior of $p(\beta)$, i.e., from the bending of the spacer.
In the case~B, the splitting of a  maximum part of the curve is less pronounced.
From the results above, it seems that ``closed'' states of the enol ring
influence the formation of the left-hand side maximum on the resulting curve in figure~\ref{fig_dipole},
or, in other words, contribute to lower values of the dipole moment compared to other conformations.

To summarize, for a curcumin molecule in MeOH and DMSO, we observe a similar behavior
of the dipole moment distribution as in water (figure~\ref{fig_dipole}).
However, the maxima at the $p(d)$ distribution are found to be a bit more pronounced
in MeOH and DMSO.
Further studies are necessary to establish the validity of these conclusions for other protic and
aprotic solvents.

\subsection{Pair distribution functions}

In order to characterize the surrounding of the curcumin molecule in different solvents,
we consider some selected  pair distribution functions (PDFs). Here,  our focus is on PDFs of solvent
molecules around the polar groups of curcumin,
since these groups formally yield the strongest attractive interactions between a curcumin molecule
and solvent species compared to other groups of a molecule. We pick up the groups where oxygen is present.
Taking into account the molecular symmetry of curcumin, some of them are combined into pairs,
e.g., for the hydroxyl groups (O5-H4 and O27-H28), we consider the functions H4/H28-OX
and O5/O27-OX (X = OW, OMe and OD)
each resulting from the averaging of the two PDFs.
A similar procedure is used for methoxy groups O2-C1 and O29-C30.
Prior to any kind of such averaging, we have checked whether the functions are similar,
up to statistical inaccuracy.

\begin{figure}[!t]
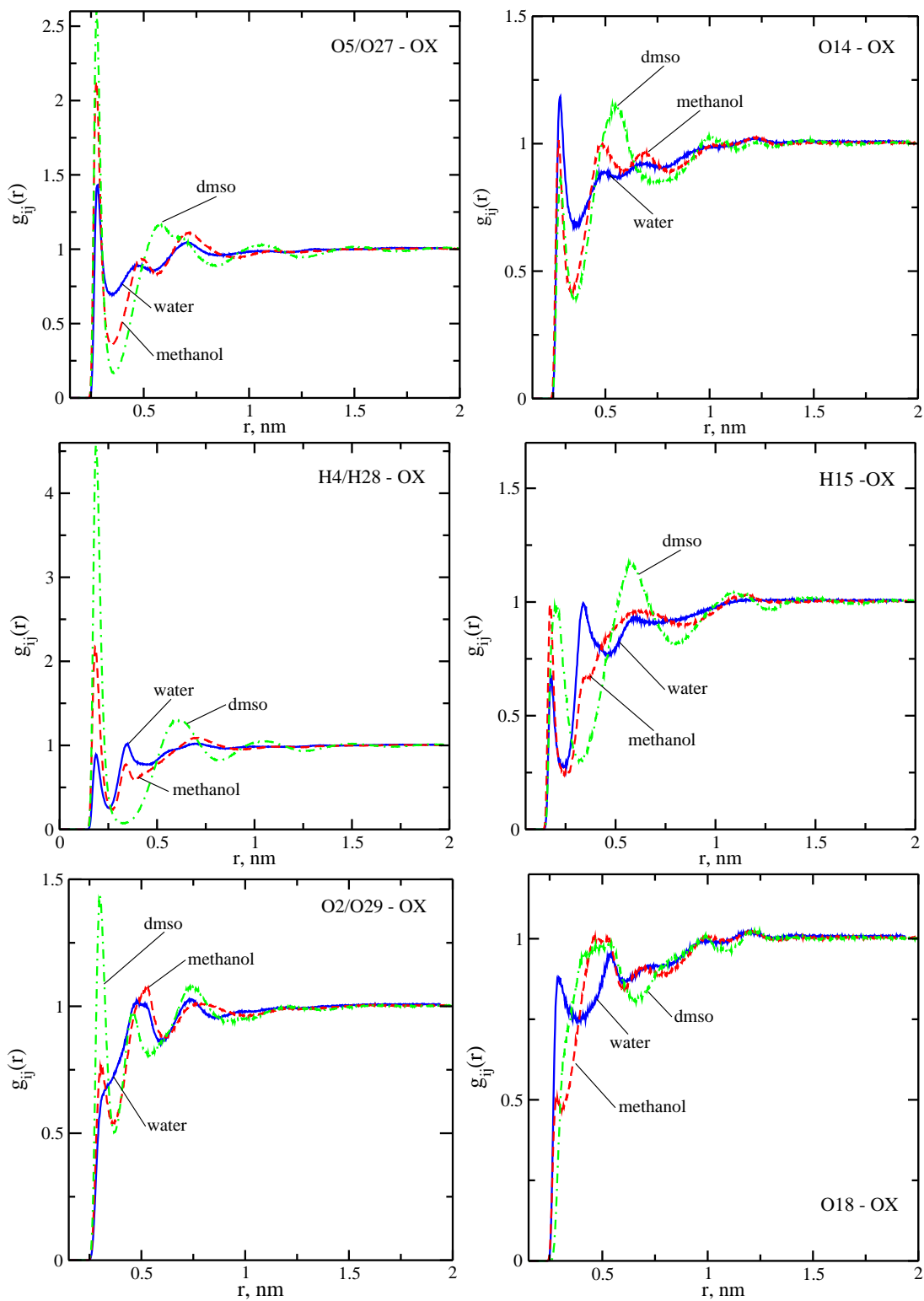

\begin{center}
\includegraphics[clip=true,height=0.45\textwidth]{rdf1}
\includegraphics[clip=true,height=0.45\textwidth]{rdf4}\\
\includegraphics[clip=true,height=0.45\textwidth]{rdf2}
\includegraphics[clip=true,height=0.45\textwidth]{rdf6}\\
\includegraphics[clip=true,height=0.45\textwidth]{rdf3}
\includegraphics[clip=true,height=0.45\textwidth]{rdf5}
\caption{\label{fig_rdf} (Color online) Pair distribution function for the hydrogens and oxygens of curcumin molecule
and oxygens (OX) of  different solvents.}
\end{center}
\end{figure}

As concerns the groups at aromatic rings, one should mention that the PDFs O5/O27-OX
are characterized by a sharp first maximum at $0.28$~nm for all solvents
in question (figure~\ref{fig_rdf}, left-hand panel).
The first maximum in MeOH and in DMSO is much higher than in water ($2.13$, $2.60$ and $1.44$, respectively).
The second peak is weakly pronounced both in water (at about $0.47$~nm) and in MeOH (at $0.49$~nm).
Actually, this is a local maximum.
Only in the case of DMSO, the second maximum is well manifested but it is not high (the height is $1.17$
at  $0.57$~nm).
The ordering of solvent species around hydrogens H4/H28  (H4/H28-OX PDFs) resembles
the above description.
Finally, for the methoxy groups, it is observed that the first peak of PDF O2/O29-OX
in the MeOH solvent is less than unity at the distance $0.30$~nm, while in water it is
absent at all. In the case of DMSO, the first maximum is well defined at the same distance $0.30$~nm,
but not as high as in the case of hydroxyl groups.

Our conclusions concerning three left-hand panels are as follows. This part of the curcumin molecule
perturbs the solvent density up to two layers at most. Distribution of solvent species around
this region of curcumin is heterogenous. The most ``dry'' part is around the methoxy group. Only
the DMSO solvent fills the first layer in the methoxy part. By contrast, the DMSO particles
form two layers in the hydroxy group region. Moreover, it seems that there exists a possibility
that OD can form hydrogen bonds with H4/H28. These trends become weaker for MeOH. Finally, in
terms of adsorption terminology, one can say that water molecules tend to their bulk rather
than approach this part of curcumin molecules or ``dewet'' it.

Now, consider the arrangement of  solvent molecules around the hydroxyl and ketone groups
at the spacer of a curcumin molecule (figure~\ref{fig_rdf}, right-hand panel).
The first maximum of PDF for O14-OX (at  $0.28$~nm) in the case
of DMSO solvent is found lower than in water and in MeOH,
for which the first maxima are observed at about the same distances.
The second maximum in DMSO
is well manifested at  $0.55$~nm and is much higher than in water and MeOH.
Those are just local maxima on a gradually increasing function that tends to unity.
For the PDFs H15-OX in water, the distribution function is similar to H4/H28-OX,
but the height of the first maximum is somewhat lower.
For  MeOH and DMSO solvents, we observe essentially lower first maxima
compared to the curves for solvent species around  hydroxyl groups at the aromatic rings.
However, the picture is different for the second maximum,
which is absent in MeOH, but well pronounced in
DMSO at  $0.58$~nm.
Finally, the PDFs for the last group on the list
of oxygen containing groups of the curcumin molecule (the ketone group) exhibits a
``desorption'' type  arrangement of solvent molecules.
The first maximum of PDF in the case of water
is less than unity at $r=0.29$~nm. In MeOH, it is much lower and in DMSO it is absent.
The second maximum in DMSO is weak. The same
is observed for the MeOH. The second maximum in water is just a local perturbation.

To summarize these observations, we would like to mention the following. The solvent density around
this part of curcumin molecule is perturbed to larger distances, cf. right-hand and left-hand three panels
of figure~\ref{fig_rdf}.
Important events occur not only at the distance of the closest approach for sites and solvent species
but in the second layer as well. Actually, the second layer should be included in the discussion
of the structure of the surroundings of a curcumin molecule. This central part of the molecules does not
exhibit much adsorptive capability, seemingly an interfacial region within this area is richer in
hydrogen bonds compared to other parts of the surface of curcumin, while in other
parts, the solvent may cover a solute without forming bonds.  These issues are touched upon herein below.

From the PDFs presented in figure~\ref{fig_rdf} we have calculated the first coordination numbers using
the definition,
\begin{equation}
 n_i(r_{\text{min}})=4\piup\rho_j\int_{0}^{r_{\text{min}}}g_{ij}(r)r^{2}\rd r
 \label{eq_nr},
\end{equation}
where $\rho_j$ is the number density of species $j$, and $r_{\text{min}}$ refers to the minimum of the
corresponding pair distribution function.
The $n(r_{\text{min}})$ indicate the quantity of oxygens of a solvent located within the first coordination shell
of the corresponding oxygen of a curcumin molecule (table~\ref{tab_coord_numb}). The distances $r_{\text{min}}$
are determined from the first minimum of the PDFs.
The coordination numbers serve as the evidence that water molecules prefer to be located in the first coordination
shells formed by oxygens of the enol ring rather than by the oxygens of the side phenol rings.
By contrast, MeOH and DMSO surround mostly the methoxy and hydroxyl groups of the side rings.
Weaker trends to observe these molecules in the first coordination shells of the enol and keto groups are
evident. Thus, the side rings play an important role in the curcumin solubility in
MeOH and DMSO solvent.

\begin{table}[!h]
\begin{center}
\caption{\label{tab_coord_numb} Position of the first minimum of the pair distribution functions and
the first coordination numbers. The sites OW, OM and OD correspond to oxygens belonging to water,
MeOH and DMSO molecules, respectively.}
\vspace{2ex}
\begin{tabular}{|l|c|c|}
  \hline\hline
  CUR-SOL  & $r_{\text{min}}$, nm  & $n(r_{\text{min}})$  \\
  \hline\hline
  O14-OW    & 0.355 & 3.17 \\
  O18-OW    & 0.390 & 4.58 \\
  O5/O27-OW & 0.350 & 3.28 \\
  O2/O29-OW     & 0.365 & 2.46 \\
  \hline
  O14-OM    & 0.336 & 0.86 \\
  O18-OM    & 0.310 & 0.35 \\
  O5/O27-OM & 0.350 & 1.54 \\
  O2/O29-OM     & 0.365 & 1.11 \\
  \hline
  O14-OD    & 0.360 & 0.58 \\
  O5/O27-OD & 0.360 & 0.97 \\
  O2/O29-OD & 0.370 & 1.00 \\
  \hline\hline
\end{tabular}
\end{center}
\end{table}

\begin{figure}[!t]
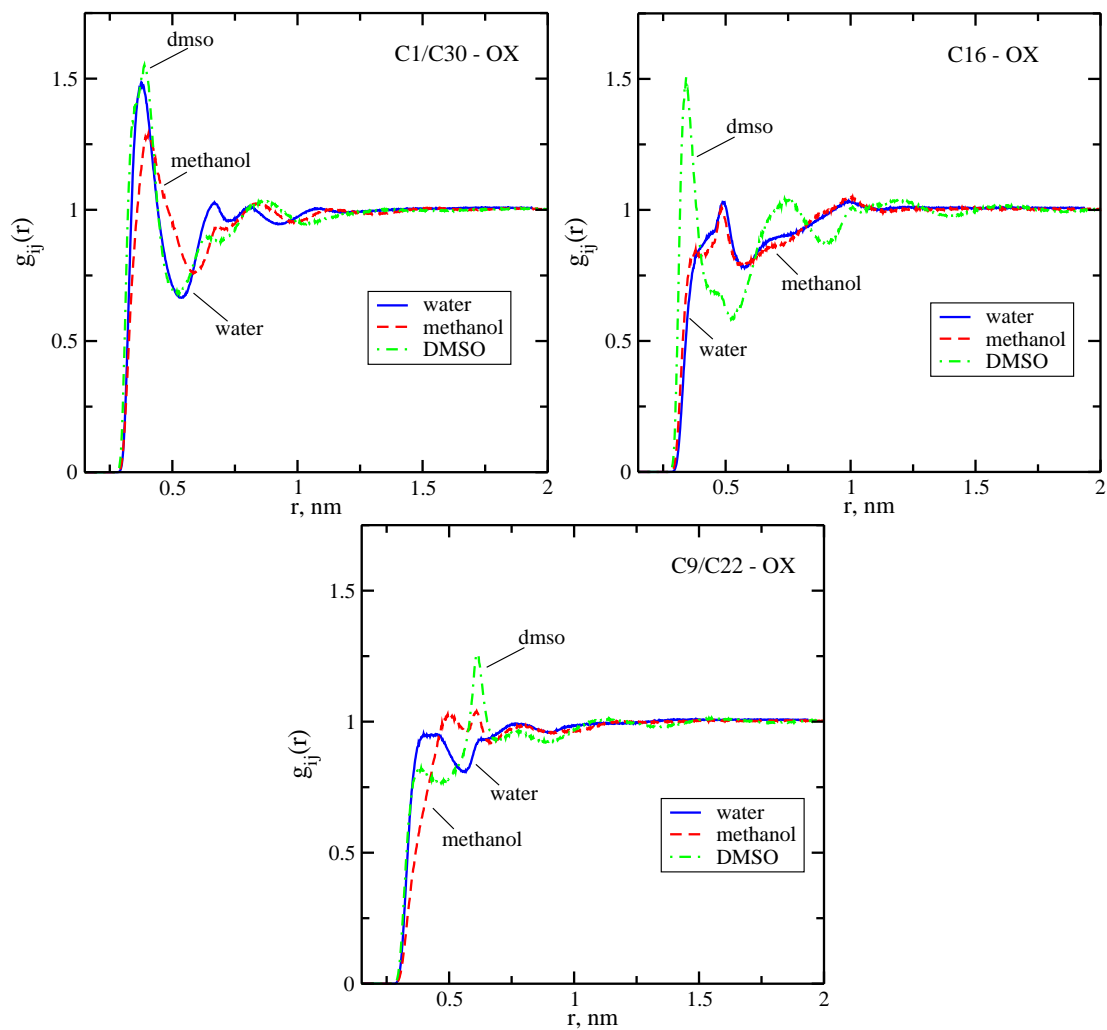

\begin{center}
\includegraphics[clip=true,height=0.45\textwidth]{rdfc1}
\includegraphics[clip=true,height=0.45\textwidth]{rdfc2}\\ 
\includegraphics[clip=true,height=0.45\textwidth]{rdfc3}
\caption{\label{fig_rdf2} (Color online) Pair distribution function for the carbons of curcumin molecule
and oxygens (OX) of different solvents.}
\end{center}
\vspace{-5mm}
\end{figure}

The arrangements of the solvent molecules around the hydrophobic groups of curcumin are analyzed
as well (figure~\ref{fig_rdf2}).
For this purpose, we consider the PDFs of solvent oxygens around several sites at the curcumin molecule,
which correspond to the sites C1/C30, C16 and C9/C22. The obtained PDFs C1-OX and C30-OX, C9-OX and C22-OX
are combined due to the symmetry of a curcumin molecule.
As expected, these PDFs have well pronounced extremal points in the case of DMSO solvent.
A less structure is observed for  water and MeOH.
However, all the solvents have PDFs indicating a rather strong augmented density close to C1 and C30 sites.

The analysis of the solvent arrangements between the two rings in the vicinity of spacer
shows that this location is preferably occupied by DMSO molecules, rather than by water or MeOH molecules.
This conclusion emerges from the well manifested  first maximum of the PDF C16-OX for DMSO at $r=0.34$~nm.
In the case of water and MeOH, the PDF C16-OX describes ``desorptive''
trends, the solvent molecules tend to the bulk rather than come close to a
curcumin molecule.
Similar trends are observed for the PDFs O9-O22.

\vspace{-1mm}
\subsection{Spatial distributions}

\begin{figure}[!b]
\begin{center}
\includegraphics[clip=true,width=0.42\textwidth]{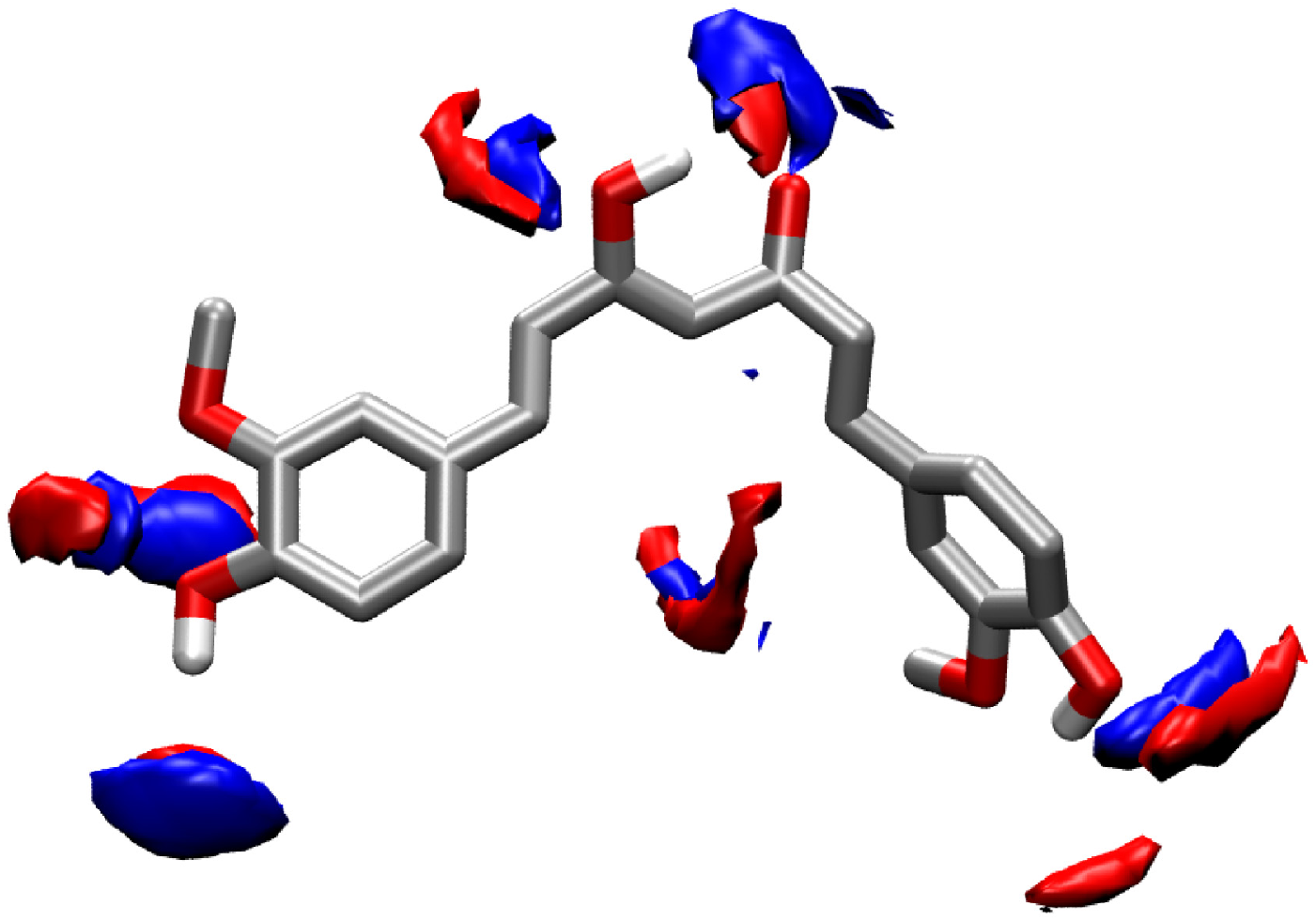}
\qquad
\includegraphics[clip=true,width=0.42\textwidth]{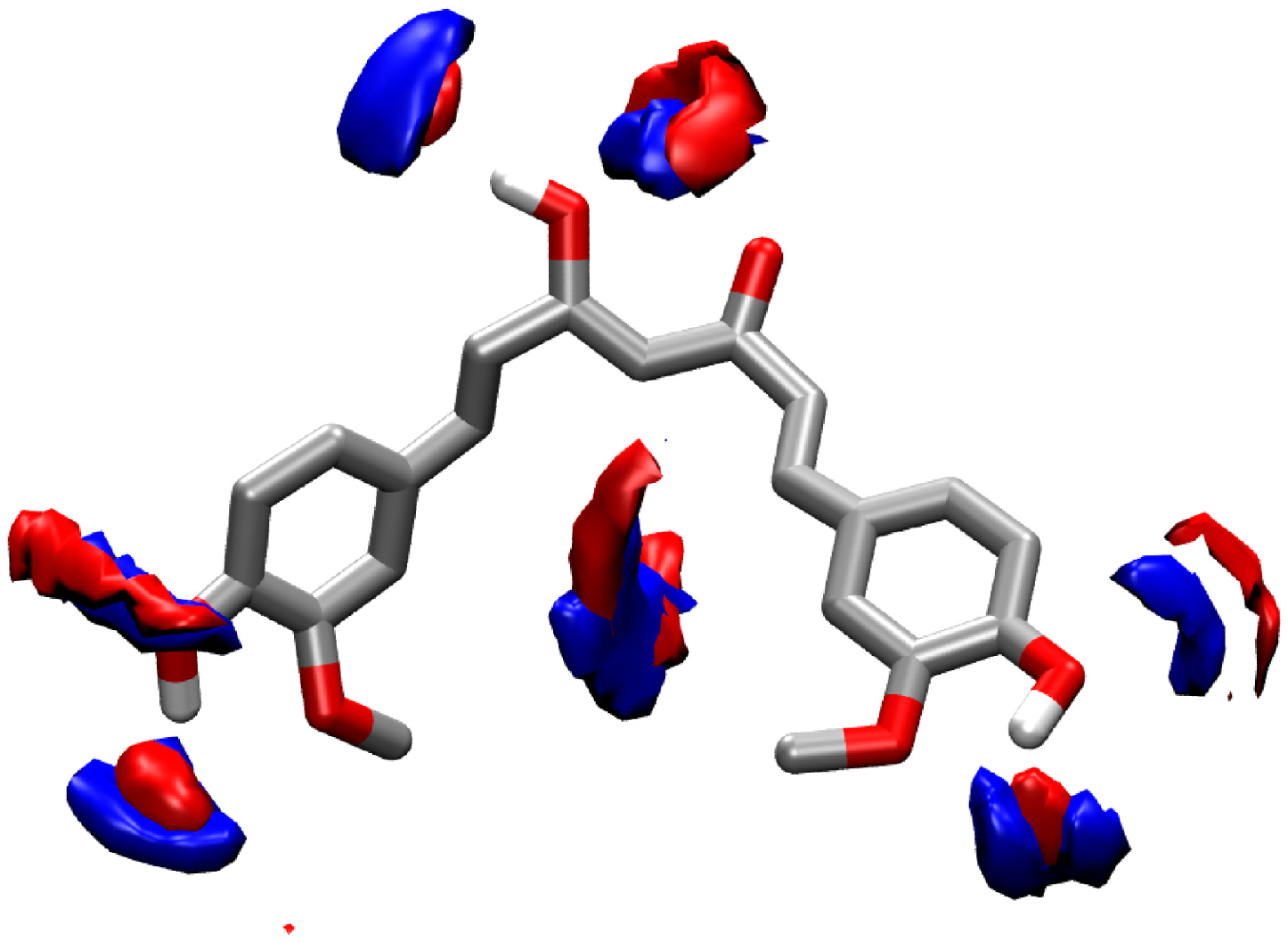}
\caption{\label{fig_spce_viz} (Color online) Spatial distribution of the density for oxygens (in red) and hydrogens (in blue)
of water molecules around a curcumin molecule for the case of the enol hydrogen
pointing toward (left-hand panel) and outward (right-hand panel) the enol group.}
\end{center}
\end{figure}
\begin{figure}[!b]
\begin{center}
\includegraphics[clip=true,width=0.42\textwidth]{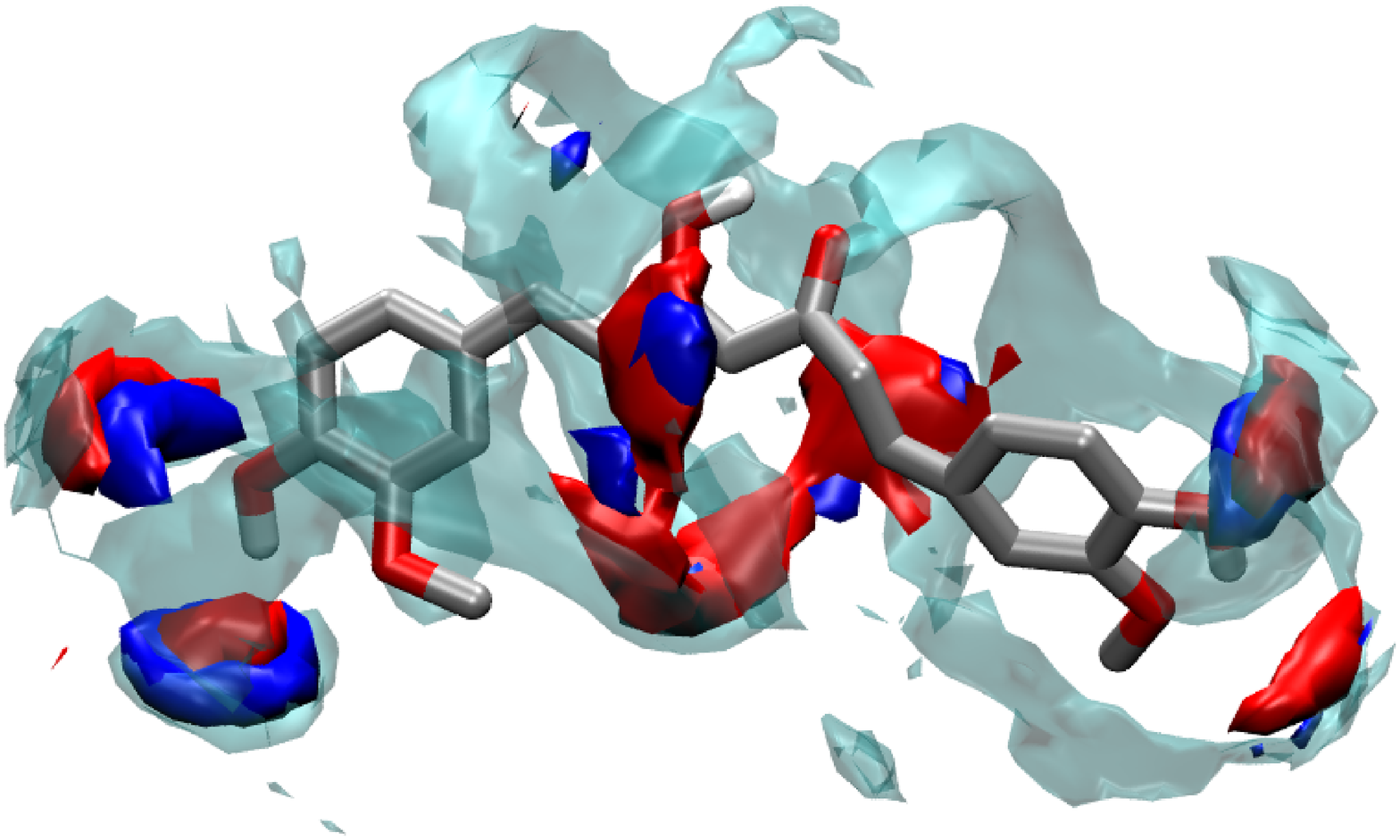}
\qquad
\includegraphics[clip=true,width=0.42\textwidth]{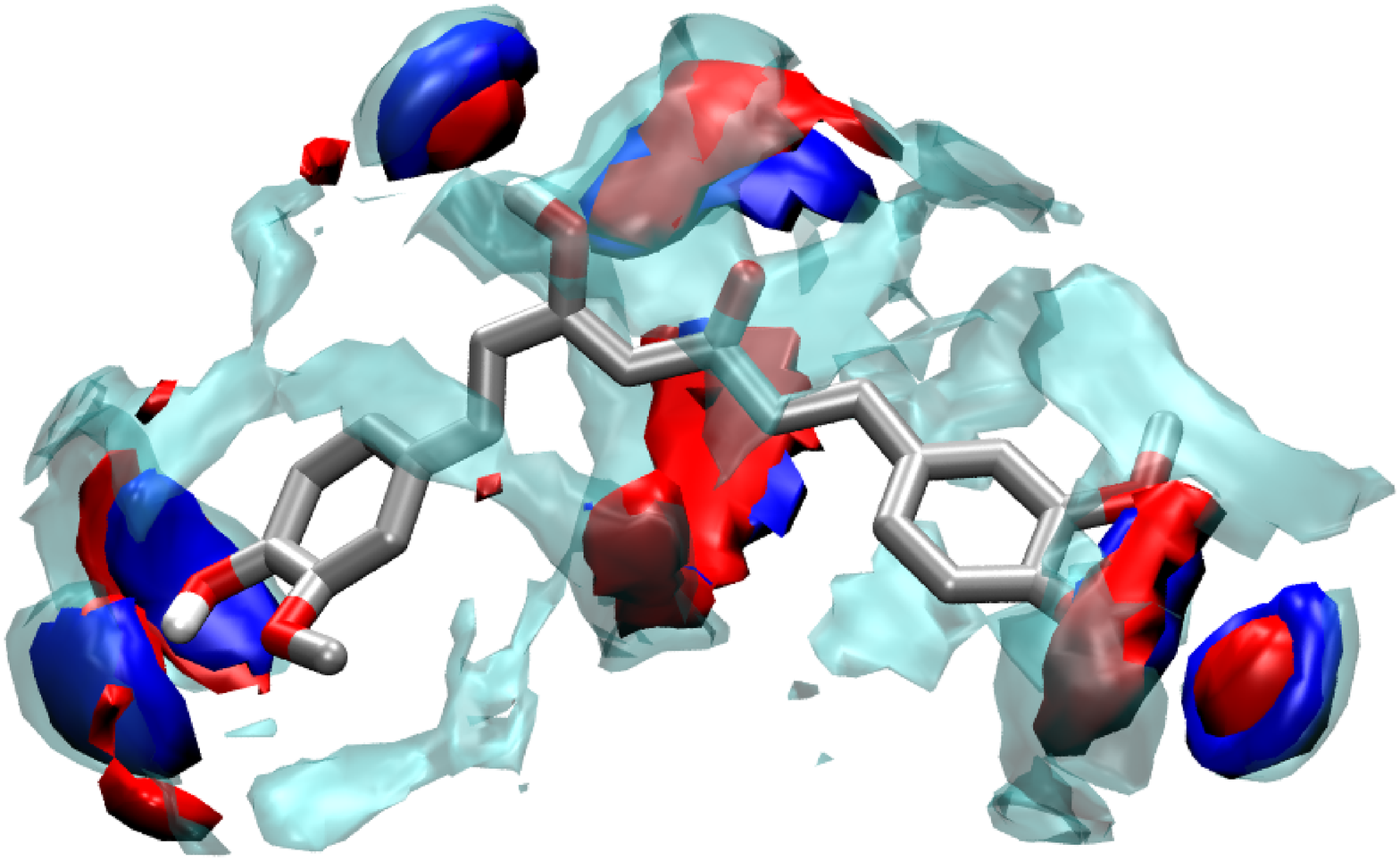}
\caption{\label{fig_meth_viz} (Color online) Spatial distribution of the density for oxygens (in red), hydrogens (in blue)
and methyl group (transparent cyan) of methanol molecules around a curcumin molecule for the case of the enol hydrogen
pointing toward (left-hand panel) and outward (right-hand panel) the enol group.}
\end{center}
\end{figure}
\begin{figure}[!b]
\begin{center}
\includegraphics[clip=true,width=0.42\textwidth]{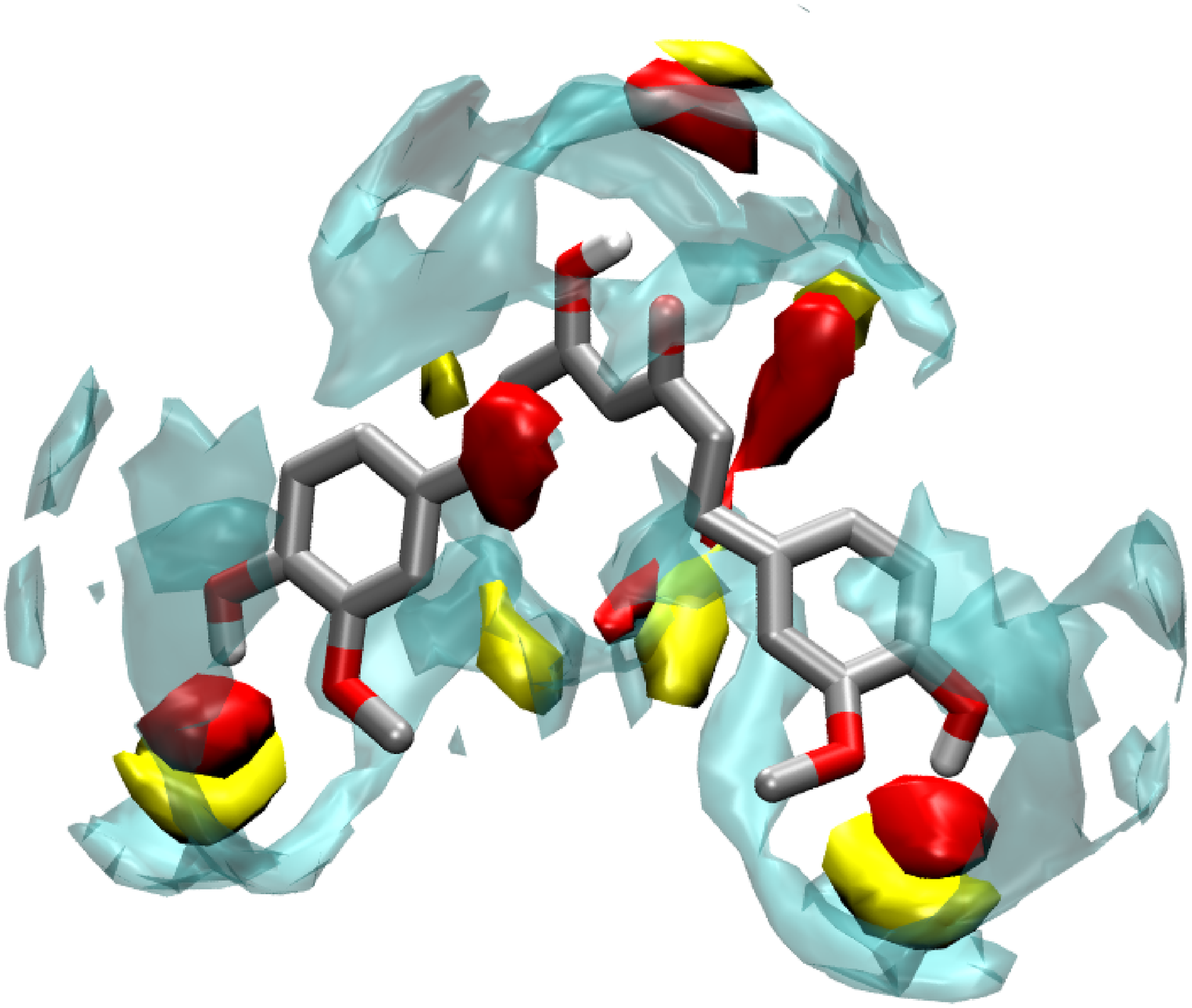}
\qquad
\includegraphics[clip=true,width=0.42\textwidth]{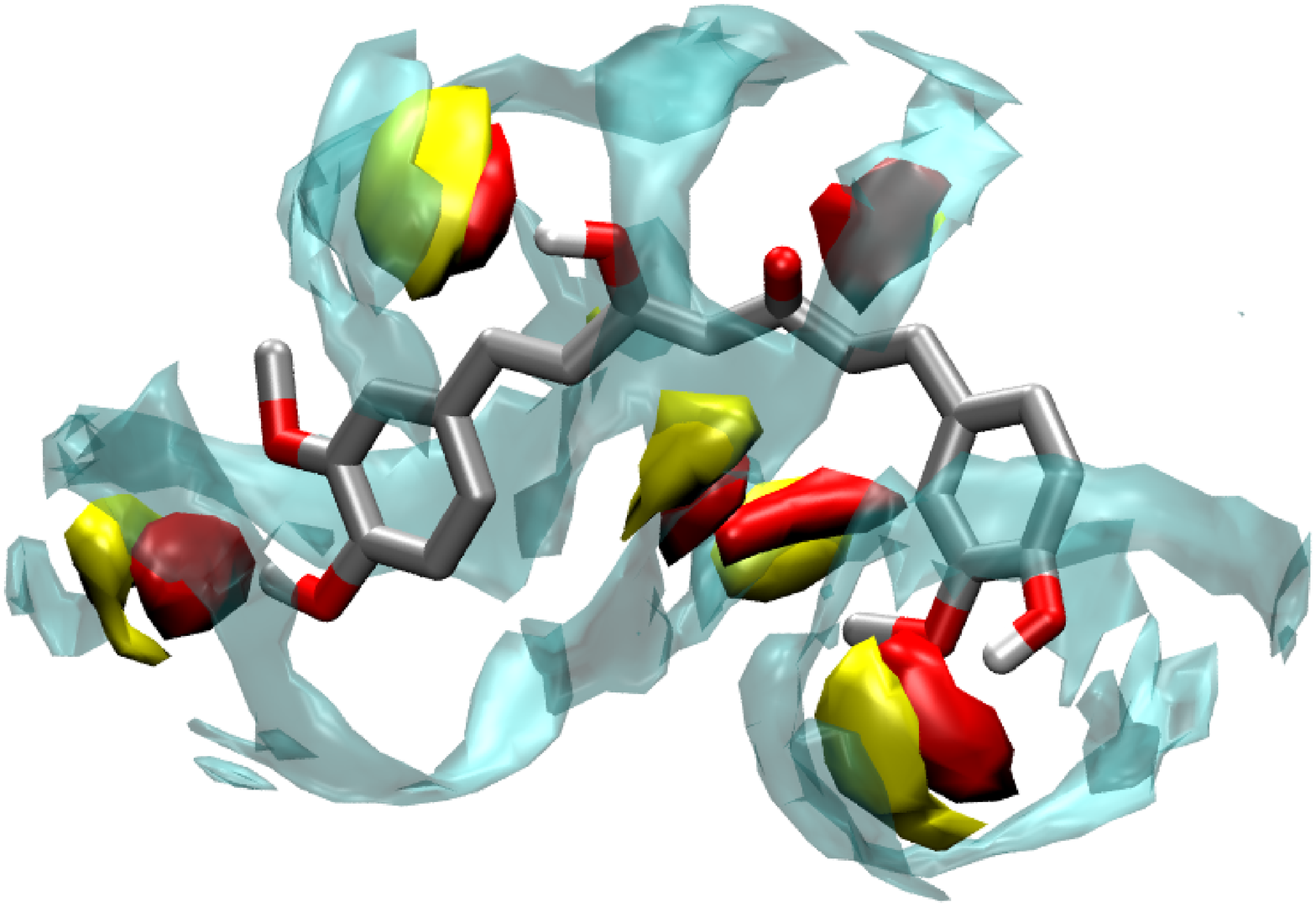}
\caption{\label{fig_dmso_viz} (Color online) Spatial distribution of the density for oxygens (in red), sulfur (in yellow)
and methyl groups (transparent cyan) of DMSO molecules around a curcumin molecule for the case of the enol hydrogen
pointing toward (left-hand panel) and outward (right-hand panel) the enol group.}
\end{center}
\end{figure}

Additional insights into the solvent surrounding the curcumin molecule
that are not accessible  from the PDFs, follow from  the analyses of the spatial density distribution
of solvent species. The corresponding plots have been built with the use of the VMD software~\cite{VMD}
in the form of isosurfaces (figures~\ref{fig_spce_viz}, \ref{fig_meth_viz} and \ref{fig_dmso_viz}).
The input data for the construction of isosurfaces of the three-dimensional density distribution function (SDF)
of solvent molecules are calculated from the simulation trajectories.
Each of the trajectories were obtained from the independent
$10$~ns runs performed in the NVT ensemble and for the fixed positions of sites of the curcumin molecule.
The values of number density for the construction of isosurfaces were chosen to make the plots
as illustrative as possible. For water oxygen, it is 0.15, whereas for methanol oxygen --- 0.045 and
finally for DMSO oxygen --- 0.050.

We have considered two specific conformations of curcumin, which correspond to the enol hydrogen pointing
toward and outward the ketone group, respectively.
In figures~\ref{fig_spce_viz}, \ref{fig_meth_viz} and \ref{fig_dmso_viz}
we discern four main regions of solvent molecules involved into curcumin surroundings:
(i)~two regions with molecules near the side rings, (ii)~a region near the enol group
and (iii)~near the spacer between the rings.
It should be noted that these regions can be split into groups. Thus, for water and MeOH,
one can observe two distinct groups near each of the side rings in the region~(i).
This can be attributed to the formation of hydrogen bonds between
the corresponding hydroxyl groups and molecules of these solvents.
A similar splitting can be seen near the hydroxyl group of enol
and the ketone group in the region~(ii).
A different situation has been found for the DMSO solvent.
Since in this case hydrogen bonding is absent,
such splitting does not occur. Moreover, it is seen that the region (ii) is affected by the enol group conformation.

In the case of water solvent, the reorientation of water molecules
in the groups of the region~(ii) is observed due to the change of enol conformation.
Finally, in all solvents, certain amount of molecules are concentrated in the region~(iii).
However, in  water and MeOH,
the molecules are grouped into belt-like cluster, while the DMSO molecules form several
distinct clusters of a rather high density,
which corresponds to the first maximum of the PDF C16-OX in figure~\ref{fig_rdf2}.

\subsection{Hydrogen bonds}
\begin{figure}[!b]
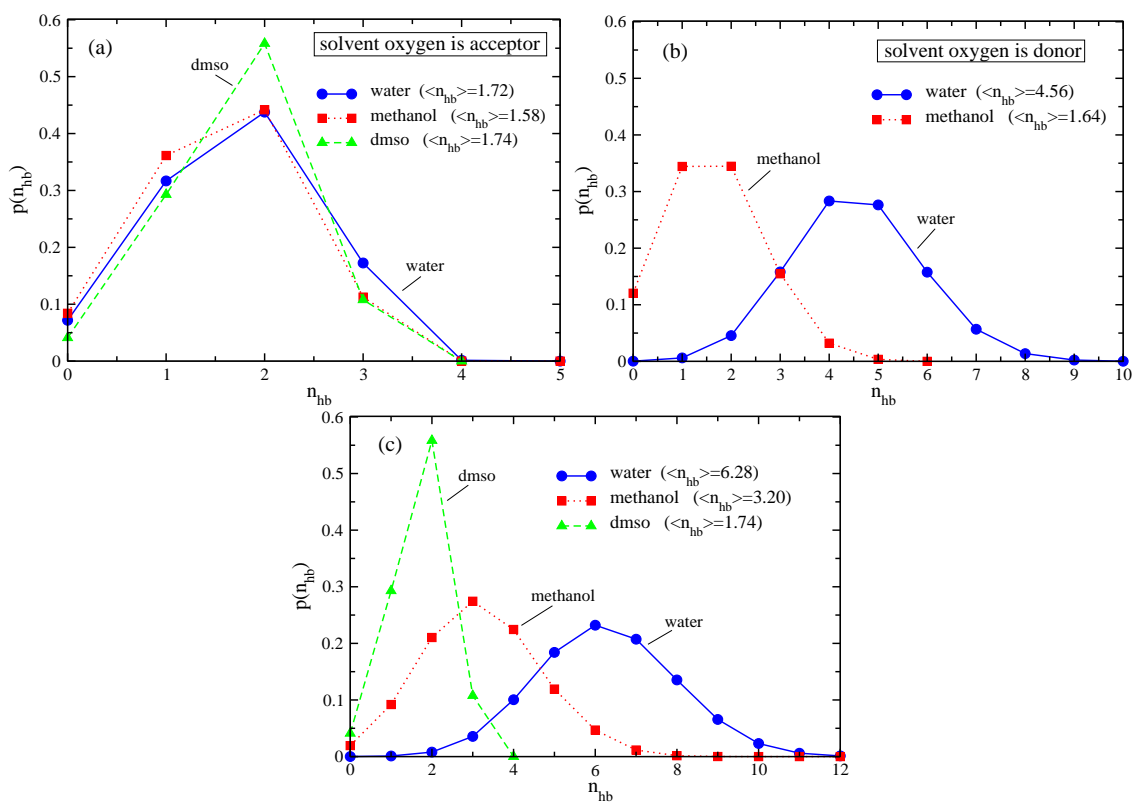

\begin{center}
\includegraphics[clip=true,width=0.49\textwidth]{hbonds_acc}
\includegraphics[clip=true,width=0.49\textwidth]{hbonds_don}
\includegraphics[clip=true,width=0.49\textwidth]{hbonds_both}
\caption{\label{fig_hbond1} (Color online) Histograms of the probability distributions of the total number of hydrogen bonds ($n_{\text{hb}}$)
between the curcumin molecule and molecules of  different solvents.}
\end{center}
\end{figure}

Solvent molecules in the vicinity of curcumin molecule are involved in the hydrogen bond~(HB) formation.
We have calculated the number of HBs, $n_{\text{hb}}$, between the oxygens or hydrogens of solvent molecules
and the groups of curcumin molecule, that contain oxygens (O14, O18, O5/O27, O2/O29). GROMACS
software was used for this purpose.
In order to identify the hydrogen bonds,  we applied a commonly used  two-parameter geometric criterium,
where donor-acceptor distance should be shorter than $0.35$~nm while the
angular cut-off is taken equal to $30$~degrees.
It is worth noting  that oxygens of water and MeOH molecules can act both as proton donors
and proton acceptors, while oxygen of DMSO can be a hydrogen bond acceptor exclusively.

In figure~\ref{fig_hbond1} we present the histograms of the probability distributions of the number of HBs.
It is observed that if an oxygen of the solvent acts as acceptor,
the distributions for all the solvents studied
are close to each other. Moreover, the maxima of these distributions obtained for water and MeOH
are practically the same, and they yield a maximum at  $n_{\text{hb}}=2$. The corresponding maximum for DMSO
is somewhat higher, but for $n_{\text{hb}}=2$ as well.
On the other hand, when oxygens of a solvent are donors,
the results are different.
Since the DMSO  oxygen cannot be a donor, the corresponding function is not present.
The distribution of HBs for water is shifted toward higher values of $n_{\text{hb}}$, and the most probable number
of HBs is in the interval $4{-}5$. This distribution has a big dispersion yielding non-zero values of probability
density even for $n_{\text{hb}}$ equal to $9$.
On the other hand, for MeOH, the distributions in figure~\ref{fig_hbond1}~(b)
are not very different from that shown in figure~\ref{fig_hbond1}~(a).
For water, the average values of $n_{\text{hb}}$ in the donor case are much larger than in the acceptor case
($4.56$ against $1.72$, respectively), while
for MeOH, the corresponding values are rather close to each other.

\begin{figure}[!b]
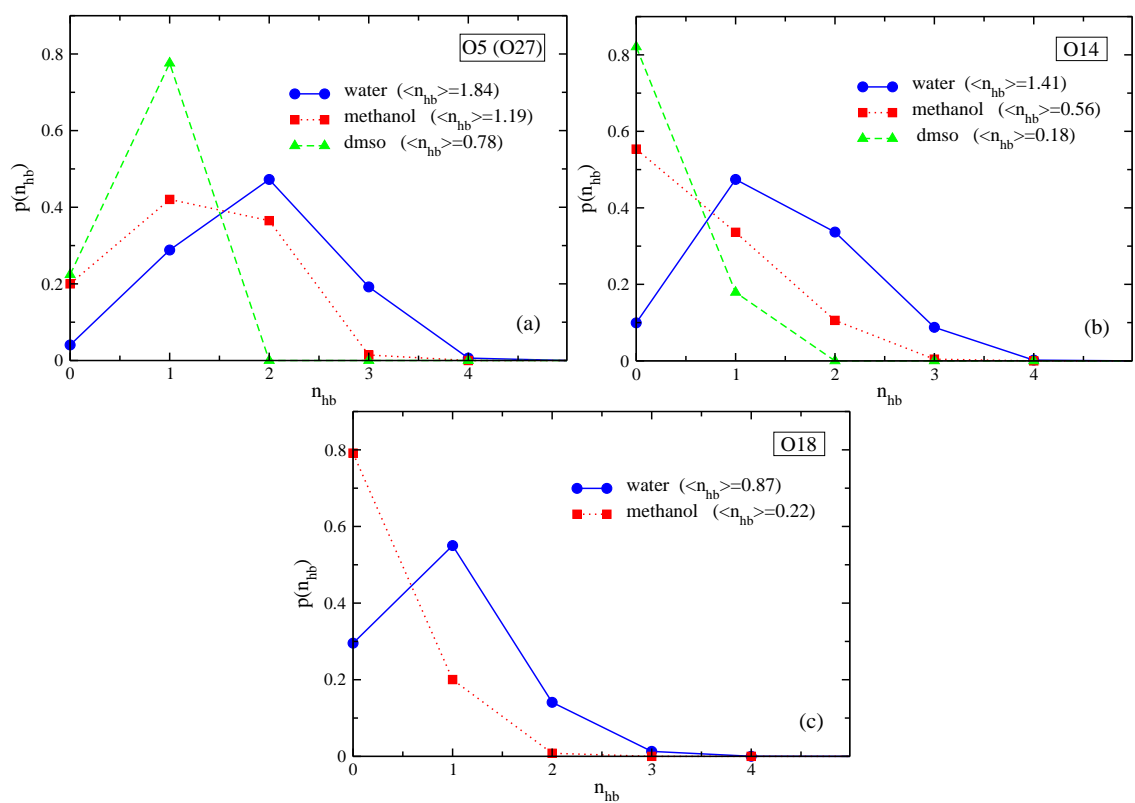

\vspace{3mm}
\begin{center}
\includegraphics[clip=true,width=0.49\textwidth]{hbonds3}
\includegraphics[clip=true,width=0.49\textwidth]{hbonds1}
\includegraphics[clip=true,width=0.49\textwidth]{hbonds2}
\caption{\label{fig_hbond2} (Color online) Histograms of the probability distributions of the number of hydrogen bonds ($n_{\text{hb}}$) formed between
the groups containing oxygens O5(O27), O14, O18  and molecules of different solvents.}
\end{center}
\end{figure}

Finally, in figure~\ref{fig_hbond1}~(c),  we present the probability distributions of the
total number of the HBs formed between solvent particles and the curcumin molecule,
regardless of the role of solvent oxygens (acceptor or donor) in hydrogen bonding.
It follows that the most probable number of HBs formed between water and curcumin is equal to $6$,
for MeOH it is $3$ and for DMSO it is equal to  $2$.
We have also calculated the averaged numbers of HBs for curcumin in each of the solvents (table~\ref{tab_hbond}).
The average number of HBs $\langle n_{\text{hb}}\rangle$ for water as solvent is  $6.28$.
It is almost twice higher than the estimate for MeOH ($\langle n_{\text{hb}}\rangle=3.20$).
At the same time, $\langle n_{\text{hb}}\rangle$ for MeOH is almost twice higher than the one calculated
for DMSO ($\langle n_{\text{hb}}\rangle=1.74$).

\begin{table}[!t]
\begin{center}
\caption{\label{tab_hbond}Average number of H-bonds between each of oxygen-containing groups
of curcumin and solvent molecules.}
\vspace{2ex}
\begin{tabular}{|l|c|c|c|c|c|c|c|} \hline\hline
& \multicolumn{3}{|c|}{water} & \multicolumn{3}{|c|}{MeOH} & DMSO \\ \hline
              & acceptor & donor & both & acceptor & donor & both & acceptor \\ \hline\hline
O14           & 0.40  & 1.01 & 1.41     & 0.25  & 0.31 & 0.56 & 0.18 \\ \hline
O18           & $-$     & 0.87 & 0.87     & $-$     & 0.22 & 0.22 & $-$    \\ \hline
O5/O27        & 0.66  & 1.18 & 1.84     & 0.66  & 0.53 & 1.19 & 0.78 \\ \hline
O2/O29        & $-$     & 0.16 & 0.16     & $-$     & 0.02 & 0.02 & $-$    \\ \hline \hline
Sum      & 1.72  & 4.56 & \bf{6.28}     & 1.58  & 1.64 & \bf{3.20} & \bf{1.74} \\ \hline\hline
\end{tabular}
\end{center}
\end{table}

In addition, we have analyzed the number of HBs formed with the particular groups of
the curcumin molecule. It is shown in figure~\ref{fig_hbond2}
that the major contribution to the total number of HBs follows from the oxygens of phenol groups (O5/O27).
For water, the most probable number of HBs occurring between the phenol groups and water is found to be equal to $2$,
for MeOH this value is in the interval $1{-}2$ whereas for DMSO it is always unity.

The probability of the formation of two HBs between distinct water molecules with the enol group (O14)
is rather high, but this is not the case for MeOH molecules that prefer to form a single bond.
For DMSO, only a single H-bond  with the enol group can be formed, although with a rather small probability.
Furthermore, hydrogen bonds can be formed between water molecules and ketone group (O18), figure~\ref{fig_hbond2}~(c).
One hydrogen bond can be formed between them with reasonable probability, the formation of two bonds being much less
probable. In DMSO, a single hydrogen bond can be formed, again with low probability.
The hydrogen bonding between solvent species and the anisole groups is quite a rare event.
The average numbers of HBs from our calculations are summarized in table~\ref{tab_hbond}.

\subsection{Self-diffusion coefficient}

\begin{figure}[!t]
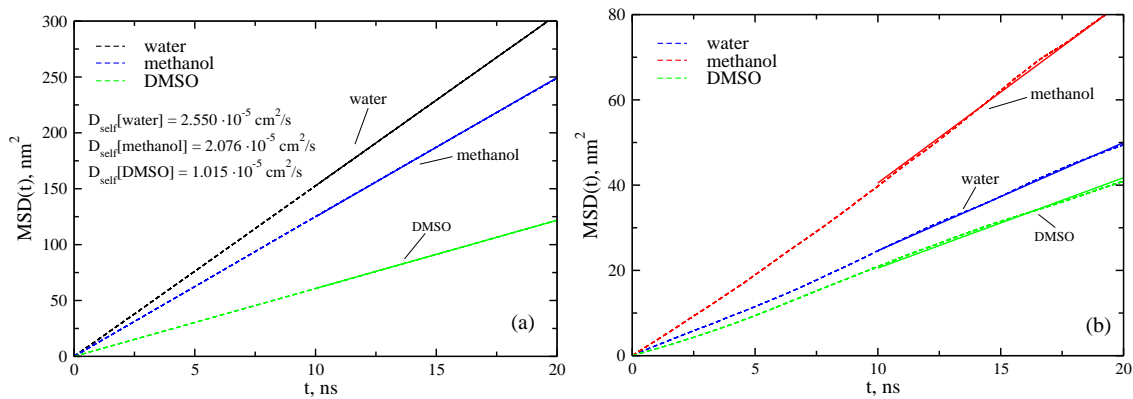

\begin{center}
\includegraphics[width=0.49\textwidth,clip]{msd_solv}
\includegraphics[width=0.49\textwidth,clip]{msd_curc}
\caption{\label{fig_msd} (Color online) Mean-square displacements of the solvents and of the curcumin molecule immersed in them.}
\end{center}
\end{figure}

The dynamic properties of a solute can be strongly affected by the solvent environment.
The self-diffusion coefficient, $D_{\text{self}}$, of a single curcumin molecule in three solvents
is estimated from the mean-square displacement (MSD) dependence on time, according to the Einstein
relation~\cite{Hansen},
\begin{equation}
D_{\text{self}} =\frac{1}{6} \lim_{t \rightarrow \infty} \frac{\rd}{\rd t} \langle\vert {\bf
r}(\tau+t)-{\bf r}(\tau)\vert ^2\rangle,
 \label{eq_selfdiff}
\end{equation}
where  $\tau$ is the time origin.
This expression is used for each species.
First, we have checked the shape of the MSDs
and the values for self-diffusion coefficients for pure solvents that
follow directly from the GROMACS software [figure~\ref{fig_msd}~(a)].
Time interval $10{-}20$~ns was used, the curves for MSD are almost linear therein.
The obtained results agree reasonably well with the already reported values
for various models of solvents considered~\cite{Gunsteren-1995,Vishnyakov-2001,Spoel-2006,Wu-2006,Hasse-2008}.

A similar procedure was then used to obtain the self-diffusion
coefficient of curcumin in each solvent [figure~\ref{fig_msd}~(b)].
The results for a single curcumin molecule follow from the trajectory of the center of mass (COM)
that can be conveniently generated by software. Moreover, the COM MSD functions on time
coming from each run (50~ns) were collected to yield  the average COM MSD curve,
which is plotted in the figure. Then,
the $D_{\text{self}}$ value was evaluated from the slope.

As one can see, the smallest value of $D_{\text{self}}$ is found for
curcumin in DMSO ($0.353\cdot10^{-5}$~cm$^{2}$/s).
In water, it is somewhat higher ($0.423\cdot10^{-5}$~cm$^{2}$/s).
A much higher value for $D_{\text{self}}$ has been obtained for curcumin in MeOH
($0.710\cdot10^{-5}$~cm$^{2}$/s).
We are not aware of the experimental data for the self-diffusion coefficient of curcumin molecule
at infinite dilution in the solvents considered.
On the other hand, the self-diffusion coefficient of curcumin in
water and methanol was reported by Samanta and Roccatano~\cite{Samanta-2013} from MD
simulation for their curcumin model.
Contrary to our results, they obtained $D_{\text{self}}$ of curcumin in water higher than in methanol
and their absolute values are much higher than ours.
This discrepancy can be due to the SPC model used in \cite{Samanta-2013} whereas
we used the SPC/E model. Additional calculations are necessary to reach a definite conclusion.

\begin{figure}[!b]
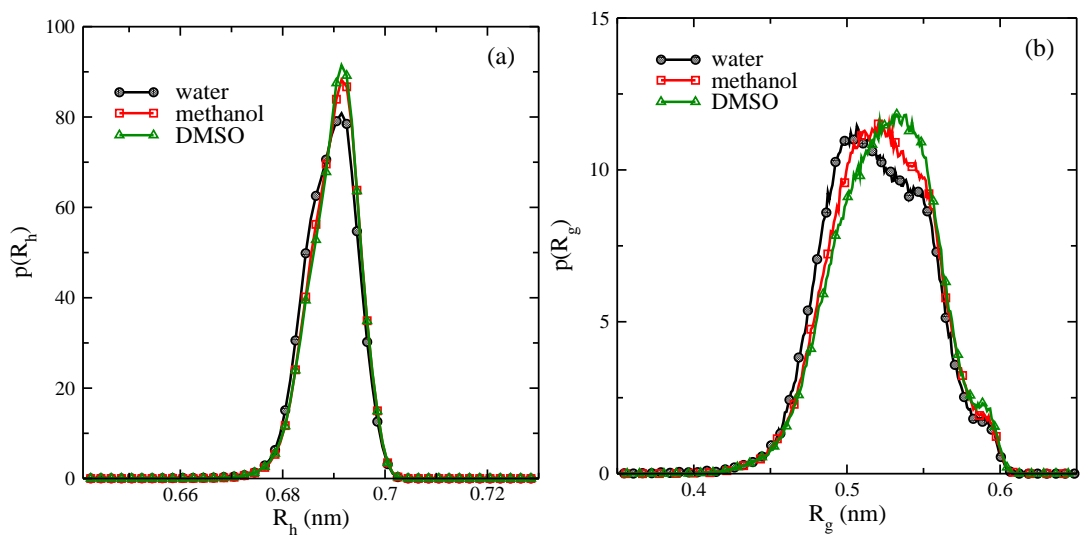

\begin{center}
\includegraphics[width=7cm,clip]{rhydro}
\includegraphics[width=7cm,clip]{gyrate}
\end{center}
\vspace{-4mm}
\caption{\label{fig_radius} (Color online) Histograms of the probability distributions of hydrodynamic radius (a) and radius of gyration (b) for the curcumin molecule in different solvents.}
\protect
\end{figure}

Nevertheless, in order to validate our results, we used the procedure based on the Stokes-Einstein equation,
\begin{equation}
 D=\frac{k_{\text B} T}{6\piup\eta_{\text S}R_{\text h}}\,,
 \label{eq_diff}
\end{equation}
where $R_{\text h}$ is the hydrodynamic radius and $\eta_{\text S}$ is the solvent viscosity.
Applying the definition,
\begin{equation}
 \frac{1}{R_{\text h}}=\frac{1}{N^{2}}\bigg\langle\sum_{i\neq j}^N \frac{1}{|\mathbf{r}_{i}-\mathbf{r}_{j}|}\bigg\rangle_{t}
 \label{eq_rhydro},
\end{equation}
we have calculated the hydrodynamic radius $R_{\text h}$ for curcumin in three solvents.
The histograms of the probability distributions of $R_{\text h}$ are shown in figure~\ref{fig_radius}~(a) and
the corresponding average values can be found in table~\ref{tab_radius}.
\begin{table}[!t]
  \centering
  \caption{\label{tab_radius}
 Hydrodynamic radius $R_{\text h}$ and the radius of gyration $R_{\text g}$ for a single curcumin molecule
 in different solvents. The experimental values of solvent viscosities $\eta_{\text S}$ were taken
 from \cite{Gonzalez} and \cite{Grande}. The self-diffusion coefficient $D$ was calculated from
 the Stokes-Einstein equation~(\ref{eq_diff}).}
 \vspace{2ex}
  \begin{tabular}{|c|c|c|c|c|c|c|}
  \hline\hline
  solvent   & $R_{\text h}$, nm & $R_{\text g}$, nm & $\eta_{\text S}$, $10^{-3}$~Pa$\cdot$s &  $D$, $10^{-5 }$~cm$^{2}$s$^{-1}$
            & $D_{\text{self}}$, $10^{-5}$~cm$^{2}$s$^{-1}$  \\
  \hline\hline
  water     & 0.689 & 0.519  & 0.890   &  0.356  & 0.423      \\
  methanol  & 0.690 & 0.523  & 0.545   &  0.580  & 0.710      \\
  DMSO      & 0.690 & 0.525  & 1.996   &  0.158  & 0.353      \\
  \hline\hline
 \end{tabular}
 \end{table}
It could be seen that the most probable values for hydrodynamic radius of curcumin
practically does not change with the solvent.
More pronounced changes are observed for the radius of gyration $R_{\text g}$,
\begin{equation}
 R^{2}_{\text g}=\frac{1}{N}\bigg\langle\sum_{i=1}^N |\mathbf{r}_{i}-\mathbf{r}_{\text{COM}}|^{2}\bigg\rangle_{t},
 \label{eq_gyrate}
\end{equation}
\begin{equation}
 \mathbf{r}_{\text{COM}}=\frac{1}{N}\sum_{j=1}^N\mathbf{r}_{j}\,,
 \label{eq_com}
\end{equation}
as it follows from the results in figure~\ref{fig_radius}~(b) given in the form of histograms of probability
distributions.
The average values for  $R_{\text g}$ are included in table~\ref{tab_radius}.
The experimental values of viscosity $\eta_{\text S}$ of water, methanol and DMSO are given
in table~\ref{tab_radius} as well. We have found that the values for the self-diffusion coefficient
calculated  from the equation~(\ref{eq_rhydro}) exhibit  similar trends upon changing the solvent
as obtained from the MSDs. However, the values differ quantitatively.
One of possible reasons is that  equation~(\ref{eq_diff}) is approximate and does not take into
account the molecular shape. However, we do not expect that the observed trends would be reversed.

\subsection{Summary and conclusions}

To conclude, in this work we have presented a very detailed description of the properties
of model solutions consisting of a single curcumin molecule in 2000 molecules of
water, MeOH and DMSO solvent. We have evaluated and analyzed the changes of the internal structure of a
curcumin molecule in terms of dihedral angles, most important characteristic distances
between particular atoms as well as the distances and angles  between segments of the molecule.
The solvent effect on the value of the dipole moment of a curcumin molecule has been
elucidated. An interface between the molecule and solvent surroundings has been characterized
in terms of pair distribution functions, coordination numbers and spatial distribution
density maps. Quantitative description of the probability of cross hydrogen bonds between
atoms of a curcumin molecule and solvent species has been given. Self-diffusion coefficient
of curcumin in three solvents has been evaluated and analyzed.

Our principal findings comprise the following points. The intrinsic bending of the molecule
already discussed in vacuum~\cite{Ilny-2016} is affected by different solvents differently.
In water, the molecule avoids contacts with solvent particles and becomes even more bent whereas
in MeOH and DMSO, more extended conformations are observed. We believe that this behavior
can be attributed to an overall hydrophobicity of the molecule. However, it would be
of interest to test this hypothesis using two or more curcumin molecules in different solvents.
 ``Switching'' of the enol hydrogen
is an important event as well, inward and outward conformations of this atom occur with
different frequency upon changing the solvent. This behavior was analyzed in terms of the
corresponding dihedral angle and histograms of the distance distribution. A similar effect
has been discussed in~\cite{Samanta-2013} within the framework of another curcumin model.
Both, the bending trends and ``switching'' of the enol hydrogen contribute to the
resulting value of the dipole moment of the molecule in solvent media. Larger values
of the magnitude of the dipole moment in comparison with vacuum~\cite{Parkanyi2004,Ilny-2016}
are observed.

The pair distribution functions built for particular atoms of a curcumin molecule
with solvent species reflect that the surroundings of the molecule are heterogenous in terms of density,
higher density is probable around phenol groups whereas ``desorption'' type effects are
more probable close to enol group. The formation of cross hydrogen bonds is most probable therein.
Seemingly, the formation of hydrogen bonds counterarrests hydrophobicity of certain fragments of
the molecule, while in water this effect is weak.
The solvent density is perturbed in approximately single layer around phenol groups whereas it is
more extended around the enol fragment. Important effects can be seen in the second layer therein.
Coordination numbers of particular atoms and density distribution maps illustrate a solvent
distribution around the molecule at a quantitative and at a qualitative level.

In the final part of this work, our results are given for the self-diffusion coefficient of a curcumin
molecule in three solvents. We presented additional arguments to prove that the trends
observed upon changing the solvent are correct by performing auxiliary calculations having
involved concepts of radius of gyration and hydrodynamic radius.

It is of much interest to extend this study along several lines.
For the moment, it seems most important to perform comparisons of the behavior of a single
curcumin molecule in other protic and aprotic solvents and extend present calculations to
a finite concentration interval for curcumin molecules in the spirit of recent work~\cite{Hazra-2014}.
Construction of bridges to experimental observations is not just desirable but certainly indispensable
for a better understanding of curcumin solutions with simultaneous improvement of the force fields.

\section*{Acknowledgements}
O.P. is grateful to D. Vazquez and M. Aguilar for technical support of this
work at the Institute of Chemistry of UNAM. Fruitful discussions with Dr. Manuel Soriano and
Dr. Hector Dominguez are gratefully acknowledged. T.P. acknowledges allocation of
computer time at the cluster of ICMP of the National Academy of Science of Ukraine and
Ukrainian National Grid.


\ukrainianpart

\title{Властивості OPLS-UA моделі однієї молекули куркуміну у воді, метанолі та диметилсульфоксиді. Результати комп’ютерного моделювання методом молекулярної динаміки}
\author{Т. Пацаган\refaddr{label1}, Я.М. Ільницький\refaddr{label1},
        О. Пізіо\refaddr{label2}}
\addresses{\addr{label1} Інститут фізики конденсованих систем НАН України, вул. Свєнціцького, 1, 79011 Львів, Україна
\addr{label2} Інститут матеріалознавства, Національний автономний університет м. Мехіко, Мехіко, Мексика
 }

\makeukrtitle

\begin{abstract}
За допомогою методу молекулярної динаміки (МД) проведено комп'ютерне моделювання та досліджено властивості однієї молекули куркуміну у воді, метанолі та диметилсульфоксиді при постійних температурі та тиску.
Для цього використано модель, запропоновану нами нещодавно для куркуміну в рамках силового поля OPLS-UA [\,J. Mol. Liq., 2016, \textbf{223}, 707] та поєднано її із моделлю води SPC/E та моделями OPLS-UA для метанолу і диметилсульфоксиду. Нами отримано детальний опис змін у внутрішній структурі розчиненої молекули, які спричинюються різними середовищами розчинника.
Проаналізовано парні функції розподілу між окремими фрагментами молекули куркуміну та молекулами розчинника. Досліджено статистичні характеристики водневих зв’язків між різними компонентами. Насамкінець, отримано коефіцієнт самодифузії молекули куркуміну в трьох модельних розчинниках.

\keywords куркумін, молекулярна динаміка, вода, метанол, диметилсульфоксид
\end{abstract}


\begin{thebibliography}{99}

\bibitem{ghosh1}
Ghosh S., Banerjee S., Sil P.C., Food Chem. Toxicol., 2015, \textbf{83}, 111, \bibdoi{10.1016/j.fct.2015.05.022}.

\bibitem{kumar1}
Kumar G.,  Mittal S.,  Sak K.,  Tuli H.S., Life Sci.,  2016, \textbf{148}, 313, \bibdoi{10.1016/j.lfs.2016.02.022}.

\bibitem{luthra1}
Luthra P.M.,  Lal N., Eur. J. Med. Chem., 2016, \textbf{109}, 23, \bibdoi{10.1016/j.ejmech.2015.11.049}.

\bibitem{mehanny1}
 Mehanny M., Hathout R.M., Geneidi A.S., Mansour S., J. Controlled Release, 2016, \textbf{225}, 1,\\ \bibdoi{10.1016/j.jconrel.2016.01.018}.

\bibitem{ngo1}
Ngo S.T., Li M.S., Mol. Simul., 2013, \textbf{39}, 279, \bibdoi{10.1080/08927022.2012.718769}.

\bibitem{nelson} Nelson K.M., Dahlin J.L., Bisson J., Graham J., Pauli G.F.,
Walters M.A., J. Med. Chem., 2017, \textbf{60}, 1620, \bibdoi{10.1021/acs.jmedchem.6b00975}.

\bibitem{wright1}
 Wright J.S., J. Mol. Struct. THEOCHEM, 2002, \textbf{591}, 207, \bibdoi{10.1016/S0166-1280(02)00242-7}.

\bibitem{ngo2}
Ngo S.T., Li M.S., J. Phys. Chem. B, 2012, \textbf{116}, 10165, \bibdoi{10.1021/jp302506a}.

\bibitem{Suhartanto-2012}
 Suhartanto H., Yanuar A., Hilman M.H., Wibisono A., Dermawan T.,
Int. J. Comput. Sci. Issues, 2012, \textbf{9}, No.~2, 90.

\bibitem{Varghese-2009}
 Varghese M.K., Molecular Dynamics Simulations of Some Nucleic Acids and their Complexes,
Ph.D. thesis, Mahatma Gandhi University, Kottayam, Kerala, India, 2009.

\bibitem{Wallace-2013}
 Wallace S.J., Kee T.W., Huang D.M., J. Phys. Chem. B, 2013, \textbf{117}, 12375, \bibdoi{10.1021/jp406125x}.

\bibitem{Samanta-2013}
 Samanta S., Roccatano D., J. Phys. Chem. B, 2013, \textbf{117}, 3250, \bibdoi{10.1021/jp309476u}.

\bibitem{Hazra-2014}
 Hazra M.K., Roy S., Bagchi B., J. Chem. Phys., 2014, \textbf{141}, 18C501, \bibdoi{10.1063/1.4895539}.

\bibitem{Sreenivasan-2014}
 Sreenivasan S., Alameen M., Krishnakumar S., Vetrivel U.,
Int. J. Pharm. Pharm. Sci., 2014, \textbf{6}, 234.

\bibitem{Yadav-2014}
 Yadav I.S., Nandekar P.P., Shrivastava S., Sangamwar A., Chaudhury A., Agarwal S.M.,
Gene, 2014, \textbf{539}, 82, \bibdoi{10.1016/j.gene.2014.01.056}.

\bibitem{Parameswari-2015}
 Parameswari A.R., Rajalakshmi G., Kumaradhas P., Chem. Biol. Interact., 2015, \textbf{225}, 21,\\ \bibdoi{10.1016/j.cbi.2014.09.011}.

\bibitem{Priyadarsini-2009}
 Priyadarsini K.I., J. Photochem. Photobiol. C, 2009, \textbf{10}, 81, \bibdoi{10.1016/j.jphotochemrev.2009.05.001}.

\bibitem{Ilny-2016}
 Ilnytskyi J., Patsahan T., Pizio O., J. Mol. Liq., 2016, \textbf{223}, 707, \bibdoi{10.1016/j.molliq.2016.08.098}.

\bibitem{Slabber-2016}
 Slabber C.A., Grimmer C.D., Robinson R.S., J. Nat. Prod., 2016, \textbf{79}, 2726, \bibdoi{10.1021/acs.jnatprod.6b00726}.

\bibitem{GROMACS}
Van der Spoel D., Lindahl E., Hess B., Groenhof G., Mark A.E., Berendsen H.J.C.,
J. Comput. Chem., 2005, \textbf{26}, 1701, \bibdoi{10.1002/jcc.20291}.

\bibitem{Kawano-2013} Kawano S.-I., Inohana Y., Hashi Y., Lin J.-M., Chin. Chem. Lett., 2013,
\textbf{24}, 685, \bibdoi{10.1016/j.cclet.2013.05.006}.

\bibitem{Kolev-2005} Kolev T.M., Velcheva E.A., Stamboliyska B.A., Spiteller M., Int. J. Quantum
Chem., 2005, \textbf{102}, 1069, \bibdoi{10.1002/qua.20469}.

\bibitem{Cornago-2008} Cornago P., Claramunt R.M., Bouissane L., Alkorta L., Elguero J.,
Tetrahedron, 2008, \textbf{64}, 8089,\\ \bibdoi{10.1016/j.tet.2008.06.065}.

\bibitem{SPCE_model}
 Berendsen H.J.C., Grigera J.R., Straatsma T.P., J. Phys. Chem., 1987, \textbf{91}, 6269, \bibdoi{10.1021/j100308a038}.

\bibitem{Leeuwen}
 Van Leeuwen M.E., Smit B., J. Phys. Chem., 1995, \textbf{99}, 1831, \doi{10.1021/j100007a006}.

\bibitem{OPLS-1996}
 Jorgensen W.L., Maxwell D.S., Tirado-Rives J., J. Am. Chem. Soc., 1996, \textbf{118}, 11225, \bibdoi{10.1021/ja9621760}.

\bibitem{Oostenbrink} Oostenbrink C., Villa A., Mark A.E., van Gunsteren W.F.,
J. Comput. Chem., 2004, \textbf{25}, 1656--1676, \bibdoi{10.1002/jcc.20090}.

\bibitem{Rosta-2009} Rosta E., Buchete N.-V., Hummer G., J. Chem. Theory Comput., 2009, \textbf{5}, 1393, \bibdoi{10.1021/ct800557h}.

\bibitem{roccatano2} Hezaveh S., Samanta S., Milano G., Roccatano D., J. Chem. Phys., 2011, \textbf{135}, 164501, \bibdoi{10.1063/1.3643417}.

\bibitem{Kurokawa-2012} Kurokawa Y., Kojima H., Yamada A., Okazaki S., Mol. Simul., 2012, \textbf{38}, 442,\\ \bibdoi{10.1080/08927022.2011.566609}.

\bibitem{VMD} Humphrey W., Dalke A., Schulten K., J. Mol. Graphics, 1996, \textbf{14}, 33, \bibdoi{10.1016/0263-7855(96)00018-5}.

\bibitem{Hansen} Hansen J.P., McDonald I.R., Theory of Simple Liquids, Academic Press, London, 2006.

\bibitem{Gunsteren-1995} Liu H., Mueller-Plathe F., van Gunsteren W.F., J. Am. Chem. Soc., 1995,
\textbf{117}, 4363, \bibdoi{10.1021/ja00120a018}.

\bibitem{Vishnyakov-2001} Vishnyakov A., Lyubartsev A.P., Laaksonen A., J. Phys. Chem. A, 2001,
\textbf{105}, 1702, \bibdoi{10.1021/jp0007336}.

\bibitem{Spoel-2006} Van der Spoel D., van Maaren P.J., J. Chem. Theory Comput., 2006,
\textbf{2}, 1, \bibdoi{10.1021/ct0502256}.

\bibitem{Wu-2006} Wu Y., Tepper H.L., Voth G.A., J. Chem. Phys., 2006,
\textbf{124}, 024503, \bibdoi{10.1063/1.2136877}.

\bibitem{Hasse-2008} Guevara-Carrion G., Nieto-Draghi C., Vrabec J., Hasse H.,
J. Phys. Chem. B, 2008, \textbf{112}, 16664, \bibdoi{10.1021/jp805584d}.

\bibitem{Gonzalez} Gonz\'{a}lez B., Calvar N., G\'{o}mez E., Dom\'{i}nguez \'{A}., J. Chem. Thermodyn., 2007,
\textbf{39}, 1578,\\ \bibdoi{10.1016/j.jct.2007.05.004}.

\bibitem{Grande} Grande M.C., Juli\'{a} J.A., Garc\'{\i}a M., Marschoff C.M., J. Chem. Thermodyn., 2007,
\textbf{39}, 1049,\\ \bibdoi{10.1016/j.jct.2006.12.012}.

\bibitem{Parkanyi2004}
 P\'{a}rk\'{a}nyi C., Stem-Beren M.R., Mart\'{i}nez O.R., Aaron J.-J.,
 Bulaceanu-MacNair M., Arrieta  A.F.,
Spectrochim. Acta, Part A, 2004, \textbf{60}, 1805, \bibdoi{10.1016/j.saa.2003.07.013}.

\end{thebibliography}
\end{document}